\documentclass[useAMS,usenatbib,usegraphicx]{mn2e}

\usepackage{amssymb}

\newcommand{\revone}{{}}
\newcommand{\revtwo}{{}}

\title[GAMA: Satellite Galaxies]{Galaxy And Mass Assembly (GAMA): the red fraction and radial distribution of satellite galaxies}

\author[Matthew~Prescott et al.]
{Matthew~Prescott,$^{1}$\thanks{E-mail:~mxp@astro.livjm.ac.uk}
I.K.~Baldry,$^{1}$
P.A.~James,$^{1}$
S.P.~Bamford,$^{2}$
J.~Bland-Hawthorn,$^{3}$
\newauthor
S.~Brough,$^{4}$
M.J.I.~Brown,$^{5}$
E.~Cameron,$^{6}$
C.J.~Conselice,$^{2}$
S.M.~Croom,$^{3}$
\newauthor
S.P.~Driver,$^{7}$
C.S.~Frenk,$^{8}$
M.~Gunawardhana,$^{3}$
D.T.~Hill,$^{7}$
A.M.~Hopkins,$^{4}$
\newauthor
D.H.~Jones,$^{4}$
L.S.~Kelvin,$^{7}$
K.~Kuijken,$^{9}$
J.~Liske,$^{10}$
J.~Loveday,$^{11}$
\newauthor
R.C.~Nichol,$^{12}$
P.~Norberg,$^{13}$
H.R.~Parkinson,$^{13}$
J.A.~Peacock,$^{13}$
S.~Phillipps,$^{14}$
\newauthor
K.A.~Pimbblet,$^{5}$
C.C.~Popescu,$^{15}$
A.S.G.~Robotham,$^{7}$
R.G.~Sharp,$^{16}$
W.J.~Sutherland,$^{17}$
\newauthor
E.N.~Taylor,$^{3}$
R.J.~Tuffs,$^{18}$
E.~van~Kampen,$^{10}$
and~D.~Wijesinghe$^{3}$
\\
$^{1}$Astrophysics Research Institute, Liverpool John Moores University,
Twelve Quays House, Egerton Wharf, Birkenhead, CH41 1LD, UK
\\
$^{2}$Centre for Astronomy and Particle Theory, University of
Nottingham, University Park, Nottingham NG7 2RD, UK
\\
$^{3}$Sydney Institute for Astronomy, School of Physics, University of
Sydney, NSW 2006, Australia
\\
$^{4}$Australian Astronomical Observatory, PO Box 296, Epping, NSW 1710, Australia
\\
$^{5}$School of Physics, Monash University, Clayton, Victoria 3800, Australia
\\
$^{6}$Department of Physics, Swiss Federal Institute of Technology
(ETH-Z{\" u}rich), 8093 Z{\" u}rich, Switzerland
\\
$^{7}$School of Physics \& Astronomy, University of St Andrews, North
Haugh, St Andrews, KY16 9SS, UK
\\
$^{8}$Institute for Computational Cosmology, Department of Physics,
Durham University, South Road, Durham DH1 3LE, UK
\\
$^{9}$Leiden University, P.O.~Box 9500, 2300 RA Leiden, The Netherlands
\\
$^{10}$European Southern Observatory, Karl-Schwarzschild-Str.~2, 85748
Garching, Germany
\\
$^{11}$Astronomy Centre, University of Sussex, Falmer, Brighton BN1 9QH, UK
\\
$^{12}$Institute of Cosmology and Gravitation (ICG), University of
Portsmouth, Dennis Sciama Building, Burnaby Road, PO1 3FX, UK
\\
$^{13}$Institute for Astronomy, University of Edinburgh, Royal
Observatory, Blackford Hill, Edinburgh EH9 3HJ, UK
\\
$^{14}$Astrophysics Group, HH Wills Physics Laboratory,
University of Bristol, Tyndall Avenue, Bristol BS8 1TL
\\
$^{15}$Jeremiah Horrocks Institute, University of Central Lancashire,
Preston PR1 2HE, UK
\\
$^{16}$Research School of Astronomy \& Astrophysics
Mount Stromlo Observatory
Cotter Road
Weston Creek, ACT 2611
Australia
\\
$^{17}$Astronomy Unit, Queen Mary University London, Mile End Rd, London
E1 4NS, UK
\\
$^{18}$Max Planck Institute for Nuclear Physics (MPIK), Saupfercheckweg
1, 69117 Heidelberg, Germany
}

\date{Accepted 2011 June by MNRAS; received in original form 2011 January}

\begin{document}               

\pagerange{\pageref{firstpage}--\pageref{lastpage}} \pubyear{2010}

\maketitle

\label{firstpage}

\begin{abstract} 

We investigate the properties of satellite galaxies that surround
isolated hosts within the redshift range $0.01 < z < 0.15$, using data
taken as part of the Galaxy And Mass Assembly survey. Making use of
isolation and satellite criteria that take into account stellar mass
estimates, we find 3\,514 isolated galaxies of which 1\,426 host a
total of 2\,998 satellites. Separating the red and blue populations of
satellites and hosts, using colour-mass diagrams, we investigate the
radial distribution of satellite galaxies and determine how the red
fraction of satellites varies as a function of satellite mass, host
mass and the projected distance from their host. Comparing the red
fraction of satellites to a control sample of small neighbours at
greater projected radii, we show that the increase in red fraction is
primarily a function of host mass. The satellite red fraction is about
0.2 higher than the control sample for hosts with $11.0 < \log_{10}
{\mathcal M}_{*} < 11.5$, while the red fractions show no difference for
hosts with $10.0 < \log_{10} {\mathcal M}_{*} < 10.5$. For the satellites
of more massive hosts the red fraction also increases as a function of
decreasing projected distance. Our results suggest that the likely
main mechanism for the quenching of star formation in satellites
hosted by isolated galaxies is strangulation.

\end{abstract}


\section{Introduction}
\label{intro}

In recent years satellite galaxies have received much attention in
both observational and theoretical studies in order to establish their
role in the formation and evolution of galaxies. In the current
$\Lambda$CDM models of the Universe, galaxies are assembled in a
hierarchical fashion, whereby small halos of dark matter (DM) merge to
form larger halos, in which baryonic matter then cools and condenses
to form stars. In this framework satellite galaxies are associated
with sub-halos of dark matter residing within the virial radii of
larger halos, which are believed to be left over from an earlier assembly
phase of their host. Thus the measurement of the spatial distribution
of satellites, both in terms of their angular and radial
distributions, can provide an insight into the mass accretion
histories of galaxies.

The most common way to determine the radial distribution is to
calculate the projected density of satellites surrounding samples of
more luminous hosts, which requires redshifts and photometry of
satellites and hosts. Previous studies which attempted to constrain
the small-scale galaxy correlation function
\citep{Lake1980,Phillips1987,Vader1991,Lorrimer1994} or investigated
the companions of field ellipticals \citep{Madore2004, Smith2004}
found that the projected density of satellites, as a function of
radius, has a profile that can be described as a power law of the form
$\Sigma(R) \propto R^{\alpha}$, with a slope, $\alpha$ ranging between
$-0.5$ and $-1.25$. Subsequently, more accurate measurements of the
projected density have been made, making use of larger and more
complete redshift surveys, such as the 2-degree Field Galaxy Redshift
Survey (2dFGRS, \citealt{Colless2001}) and the Sloan Digital Sky Survey
(SDSS, \citealt{York2000}), allowing the production of large statistical
samples of satellites and hosts selected using various well defined
criteria \citep{Sales2004,Chen2006, Ann2008,Chen2008, Bailin2008}.

Studies which have examined the radial distribution of satellites as a
function of host luminosity, colour or morphology, have obtained mixed
results. The first to divide their sample into early and late types
was \cite{Lorrimer1994}, who found the projected density profile of
the satellites of early-type hosts to have steeper slopes and
therefore to be more centrally concentrated than the satellites of
late-type hosts. In contrast to this, \cite{Sales2005} using data from
2dFGRS, found that the distribution of the satellites of red hosts has
a shallower profile than the satellites of blue hosts, which even
deviates from a power law, flattening at small projected
separation. More recently \cite{Chen2008} using SDSS data, found that
the satellites of both red and blue hosts follow similar power slopes after
correcting for interlopers (galaxies mistaken as satellites through
projection but not actually physically bound to their host).

Dividing their satellite sample into red and blue populations,
\cite{Chen2008} also finds that red satellites are more centrally
concentrated than blue satellites, a trend which is also seen in the
semi-analytic galaxy samples produced by \cite{Sales2007} using the
Millennium Simulation. One explanation of this is that red satellites
were accreted into their host's halo at earlier times than the blue
satellites, which is consistent with the observational finding that
red satellites have an anisotropic angular distribution with a
preference of being aligned along the major axes of their hosts
\citep{Brainerd2005,
  Yang2006,Azzaro2007,Bailin2008,Agustsson2010}.
Comparisons between the radial distribution of satellites and the dark
matter distribution produced from simulations have also been conducted
by \cite{Chen2006} and \cite{VDBosch2005}, who find that satellites are more
centrally concentrated than dark matter sub-halos, but
consistent with the dark matter profile.

Satellite galaxies not only provide useful information about the
formation of galaxies but also about the processes that govern galaxy
evolution in localised environments on scales of $\sim 1$\,Mpc. In the
current theory of galaxy evolution it is thought that virtually all
galaxies start off as blue, late-type discs which are then
transformed by various processes into red, early types. This
is supported by the observed bimodality of galaxies which can be seen
out to $z \sim 1$ \citep{Bell2004, Willmer2006, Prescott2009}, and
studies such as \cite{Willmer2006} and \cite{Faber2007} which have shown there
has been a doubling in the stellar mass density of galaxies on the red
sequence over the last 7-8 Gyr.

The main processes believed to be responsible for transforming
blue/late-type galaxies into red/early types involve the quenching of
star formation, and include major-mergers \citep{Toomre1972,
  Hopkins2008a}, feedback from active galactic nuclei
\citep{Bower2006,Croton2006} and the depletion of gas reservoirs that
fuel star formation. For satellite galaxies, the dominant process is
most likely to be gas depletion, caused by the stripping of gas via a
number of different hydrodynamical and radiative interactions with
their hosts, acting over different timescales.

{\revone When a satellite halo is accreted by the larger halo of its host, hot
  gas from the satellite may be removed in the process known as strangulation
  \citep{Larson1980, Balogh2000}, resulting in the gradual decline in star
  formation over long timescales ($> 1$ Gyr), as its fuel for future star
  formation is depleted. Star formation can be shut off more rapidly if the
  satellite is subjected to sufficient external pressure that its cold gas
  reservoir is removed in the process of ram-pressure stripping
  \citep{Gunn1972}. Gas stripping via harassment \citep{Moore1996} whereby the
  dark matter sub-halos of satellites are heated after undergoing frequent
  high-velocity encounters with other dark matter halos, is also a
  possibility, although this process is more likely to occur in galaxy
  clusters rather than small groups.

Recent studies using the SDSS have indicated that strangulation is the
main mechanism causing the transition of satellites from the blue to
the red sequence. Producing a group catalogue from DR2,
\cite{Weinmann2006} investigate how the fractions of early- and
late-type satellites vary as a function of halocentric radius, halo
mass and luminosity, {\revtwo observing that the early-type fraction 
increases with decreasing halocentric radius, increasing halo mass and 
increasing luminosity. They argue that the increase in early-type fraction
with luminosity at fixed halo mass is not expected if ram-pressure 
stripping or harassment is the primary cause of gas removal.}
\cite{Ann2008} use SDSS DR5 to investigate how the early-type
fractions of satellites surrounding isolated hosts vary as a function
of luminosity and projected distance. 
They find that the early-type satellite fraction increases significantly with
decreasing projected radius for early-type hosts and stays
approximately constant for late types. They conclude that hot X-ray
emitting gas of the early-type hosts is responsible for the {\revtwo removal}
of gas.

Using an SDSS DR4 group catalogue, \cite{VDBosch2008} compare the
concentrations and colours of centrals and satellites of the same
mass. By matching central and satellite pairs in both stellar mass and
concentration, they find there is a significant difference in
colour. Centrals and satellites matched in both stellar mass and
colour, on the other hand, show no difference in concentration. Under
the assumption that centrals are the progenitors of satellites
(centrals change into satellites after being accreted into a larger
halo), this implies that either the strangulation or ram pressure
stripping process is occurring rather than harassment, which is believed to
have a significant effect on the morphology of galaxies. Investigating
the red fraction of satellites by mass they estimate that 70 per cent
of satellite galaxies with ${\mathcal M}_{*} = 10^{9} {\mathcal
  M}_{\odot}$ have undergone satellite quenching in order to be on the red
sequence at the present, with 30 per cent already red before becoming
a satellite. For more massive satellites they find quenching to be
less effective, with 65 per cent of satellites with ${\mathcal M}_{*}
= 10^{10} {\mathcal M}_{\odot}$ being red before accretion and
virtually all satellites with ${\mathcal M}_{*} = 10^{11} {\mathcal
  M}_{\odot}$ already being red before becoming satellites.}

Furthermore, satellites are believed to affect the evolution of
their hosts. Minor mergers between dwarf satellites and their central
hosts provide one way in which to distort and thicken galaxy discs
\citep{Quinn1993} and enlarge the bulge components of discs
\citep{Dominguez2008}, and mergers involving gas-rich satellites may
also provide gas which could prolong or replenish star formation in
early-type spirals \citep{White1991,Hau2008}.
 
{\revone Finally, satellite systems analogous to the Milky Way
  (MW)-Magellanic clouds systems have also become of much interest
  lately due to their apparent rarity. Observational studies such as
  \cite{James2011}, \cite{Liu2011} and \cite{Tollerud2011} find that
  only $\sim$10 per cent of MW like hosts have one Magellanic cloud
  like satellite and only $\sim$5 per cent have two. Similarly N-body
  simulations have shown that less than 10 per cent of MW sized dark
  matter haloes contain two Magellanic cloud sized sub-halos
  \citep{Busha2010,Boylan-Kolchin2010}.}

In this paper we use data taken from the Galaxy And Mass Assembly
(GAMA) survey to investigate: firstly the radial distribution of
satellites surrounding a sample of isolated host galaxies, as a
function of host mass and colour; and secondly how the red fraction of
satellites depends on the projected distance between satellite and
host, the stellar mass of the satellite and the stellar mass of the
host. {\revtwo The GAMA redshift survey extends up to 2\,mag deeper
  than the SDSS Main Galaxy Sample and, unlike previous studies, we
  use isolation criteria that take account of stellar mass
  estimates. The results are discussed in terms of} the quenching
mechanisms that act on satellites.

The structure of this paper is as follows. In Section~2 we outline the
GAMA survey from which our samples of galaxies are taken. In Section~3
we define our criteria used to select isolated hosts and satellite
galaxies. We also define how we select a control sample of
neighbours. Section~4 describes how we divide the red and blue
populations of galaxies, compare properties of the hosts and
satellites, and determine the projected density of satellites as a
function of host mass and host colour. We show how the red
fraction of satellites depends on satellite mass, host mass and
projected radius and discuss potential processes that could
produce these results in Section~5. Finally in Section~6 we summarise
our main results. Throughout this paper we assume values of $H_0 = 70$
km\,s$^{-1}$ Mpc$^{-1}$, $\Omega_{m} = 0.3$ and $\Omega_{\Lambda} =
0.7$.
                  
\section{Data}

\subsection{Galaxy And Matter Assembly} 

GAMA is a project to construct a multi-wavelength (far-UV to radio)
database of $\sim 375\,000$ galaxies, by combining photometry and
spectroscopy from the latest wide-field survey facilities
\citep{Driver2009, Driver2011}. Currently covering
$144$ deg$^{2}$ and going out to $z \sim 0.5$, GAMA will allow the
study of galaxies and cosmology on scales between 1~kpc and 1~Mpc and
provide the link between wide-shallow surveys, such as the SDSS Main
Galaxy Sample \citep{Strauss2002}, 2dFGRS 
\citep{Colless2001} and 6dFGRS \citep{Jones2004}, and 
narrow-deep surveys DEEP2 \citep{Davis2003} and VVDS \citep{LeFevre2005}.

Central to GAMA is a redshift survey conducted at the 3.9-m
Anglo-Australian Telescope (AAT) using the AAOmega spectrograph
\citep{Sharp2006}, which is crucial to addressing the main objectives
of the project. These include determining the dark matter halo mass
function of groups and clusters \citep{Eke2006}, measuring the stellar
mass function \citep{Baldry2008} of galaxies down to Magellanic Cloud
masses and determining the recent galaxy merger rate
\citep{DePropris2005}.  

The redshifts used in this paper were obtained as part of the initial
spectroscopic survey known as GAMA I, carried out over 66 nights
between March 2008 and May 2010. This consists of three $12 \times 4$
deg fields at 9h, 12h and 14.5h (G09, G12 and G15) along the
celestial equator and covering in total $144$ deg$^{2}$. For detailed
descriptions of the spectroscopic target selection and the tiling
strategy used for GAMA, the reader is referred to \cite{Baldry2010}
and \cite{Robotham2010}, respectively. In brief, galaxies are selected
for spectroscopy using an input catalogue drawn from the Sloan Digital
Sky Survey (SDSS) Data Release 6 \citep{SDSSDR6} and UKIRT Infrared
Deep Sky Survey (UKIDSS) \citep{Lawrence2007}. 

In the following analysis we use data for galaxies which make up the
{\it r}-band limited Main Survey, which contains in total 114\,441
galaxies, which are spectroscopically selected to have Galactic
extinction corrected Petrosian magnitudes \citep{Petrosian1976} of
$r_{Petro} < 19.4$ in fields G09 and G15 and $r_{Petro} < 19.8$ in G12.

The high density of spectra per square degree of sky and high completeness of
the redshift survey (98 per cent down to $r_{Petro} = 19.8$,
\citealt{Driver2011}) required to achieve the main objectives of the project,
make GAMA an ideal dataset to study satellite galaxies, since it does not
suffer from the same incompleteness as other spectroscopic surveys due to
fibre collisions. The SDSS for example has a minimum fiber spacing of 55
arcsec, resulting in 10 per cent of SDSS targets being missed from the
spectroscopic sample {\revtwo because each area is generally tiled only once,
  or twice in the overlap between plates.  The fraction missed is higher in
  higher density regions, e.g., for galaxies with two other targets within 55
  arcsec, the chance of obtaining an SDSS redshift is half that of galaxies
  with no close neighbours (figure~3 of \citealt{Baldry2006}).  This is not
  true for GAMA because each area is tiled 4 or more times, close targets are
  given a higher priority in early visits, and high completeness is a primary
  goal \citep{Robotham2010}.}

\subsection{Distances}

From the 114\,441 galaxies in the {\it r}-band magnitude limited Main Survey we
choose galaxies with reliable redshifts (redshift quality values of $Q
\ge 3$), in the range $0.01 < z < 0.15$, which results in a sample of
34\,102 galaxies, from which we search for isolated galaxies and
satellites. In the low redshift regime of this galaxy sample, the
recessional velocities of galaxies are significantly affected by
peculiar motions which can cause distance estimates to be in error, if
simply assuming Hubble flow velocities \citep{Masters2004}. To
mitigate the effects of peculiar motions on distance we make use of a
parametric model of the local velocity field known as the
multi-attractor model of \cite{Tonry2000}.
 
\subsection{Photometry}
 {\revone The photometry used in this paper includes {\it gri} Petrosian
   magnitudes taken from SDSS DR6 and Kron-like AUTO magnitudes
   measured using SExtractor \citep{Bertin1996}, from our own
   re-reduction of the SDSS images described in
   \cite{Hill2011}}. In brief the photometry for the SDSS is obtained
 for five broad-band filters ({\it ugriz}) using a dedicated 2.5-m
 telescope at Apache Point, New Mexico, equipped with a mosaic CCD
 camera \citep{Gunn1998} and calibrated with a 0.5-m telescope
 \citep{Hogg2001}. For greater detail regarding the SDSS the reader is
 referred to \cite{York2000} and \cite{Stoughton2002}.

Unless otherwise stated, absolute magnitudes are calculated from GAMA AUTO
mags such that:
\begin{equation}
M = m_{AUTO} - 5\log_{10}{D_{L}}- 25 - A - K 
\end{equation}
where $m_{AUTO}$ is the apparent AUTO magnitude of a galaxy, $D_{L}$
is the luminosity distance in Mpc, $A$ is the Galactic extinction from
\cite{Schlegel1998} and $K$ the K-correction to z = 0.0, determined
using the K-Correct v\_4\_1\_4 code of \cite{Blanton2007}. For the
$\sim 700$ galaxies where SExtractor has failed to assign an AUTO mag
in any of the {\it g,r,i}-bands we use the Petrosian magnitude
($m_{Petro}$) to calculate absolute magnitudes.

In the next sections we describe the criteria and methods used in the
search for isolated and satellite galaxies.

\section{Selection of Satellite Systems}

\subsection{Isolated Galaxies}

Before searching for satellites we must define a set of criteria to
search for isolated galaxies which may host them. In previous papers
isolated galaxies are usually found by searching for galaxies which
have no neighbours brighter than a given magnitude (either apparent or
absolute) contained within a cylinder, determined by a
velocity and projected radius
\citep{Zaritsky1993,Sales2004,Bailin2008,Ann2008}.
Instead of using an isolation criteria that depends solely on
 luminosity, here we use a criterion that also takes into account an
 estimate of the galaxies' stellar masses, which makes the selection
 more physically motivated.

To determine stellar masses we use an expression based on the relationship
between a galaxy's $(g-i)$ colour and $i$ band stellar mass-to-light ratio
(${\mathcal M}_{*}/L$) from \cite{Taylor2011}, which is given by:
\begin{equation}
  {\log}_{10}{\mathcal M}_{*} = -0.68 + 0.73 (g-i) - 0.4 (M_{i} - 4.58)  
\end{equation}  
Here ${\mathcal M}_{*}$ is the stellar mass of the galaxy in solar units
(${\mathcal M}_{*}/{\mathcal M}_{\rm solar}$), $(g - i)$ is the rest-frame
colour, and $M_{i}$ the absolute $i$-band magnitude of the galaxy calculated
using GAMA AUTO magnitudes.  
This way of estimating stellar masses has the advantage of only using two
luminosities in a transparent way, unlike estimating stellar masses determined
via SED fitting; {\revtwo and several authors have recently suggested that
${\mathcal M}_{*}/L$ correlates most reliably with $g-i$
\citep{Gallazzi2009,Zibetti2009,Taylor2010}.}

{\revone The central 95 per cent of the rest-frame $(g-i)$ colour distribution
  of the galaxy sample considered in this study corresponds to $0.26 < (g-i) <
  1.24$, {\revtwo implying that the galaxy stellar mass-to-light ratio in the
    $i$-band typically varies by as much as a factor of 5 from 0.32 to
    1.68. In order to be robust against colour errors we restrict the
    mass-to-light ratio to this range.}}

In the search for isolated centrals we limit ourselves to using galaxies with
$r_{Petro} \le 19.4$ in GAMA fields G09 and G15, and $r_{Petro} \le 19.8$ in
GAMA field G12.  In order to be considered isolated, the centrals must not
have any comparably massive neighbours within a large surrounding region. We
define isolated galaxies as those which which have:

\begin{enumerate}

\item A stellar mass more than 3 times that of any
  neighbours $({\mathcal M}_{*,Iso} > 3{\mathcal M}_{*,Neighbour})$,
  within a projected radius of $R_{p} \le 1$\,Mpc and $|c \Delta z| \le
  500$ km\,s$^{-1}$.

\item An apparent {\it r}-band Petrosian magnitude such that:
\begin{equation}
 r_{Petro} < r_{lim} - 2.5(\log_{10}\, 3) - \Delta r
\end{equation}
 with $r_{lim}$ being 19.4 for G09 and G12 or 19.8 for G15 and where
 $\Delta r$ is given by:
\begin{equation}
 \Delta r = 2.5 [ \log_{10}({\mathcal M}_{*}/L_{r})_{max} - 
                  \log_{10}({\mathcal M}_{*}/L_{r}) ]
\end{equation} 
and $({\mathcal M}_{*}/L_{r})_{max}$ is a `maximum' mass-to-light ratio in
the {\it r}-band. {\it r}-band mass to light ratios are determined by
dividing the stellar masses by absolute {\it r}-band Petrosian
luminosities in solar units. This maximum stellar mass ratio is chosen
to have the value $({\mathcal M}_{*}/L_{r})_{max} = 2.14$. The purpose of
this condition is that it ensures that isolated galaxies are
sufficiently brighter than the limit of the field, such that even
neighbouring galaxies with the highest mass-to-light ratios that are
fainter than the limit still have a stellar mass that is a factor 3
less than that of the central.

\item {\revone A projected distance greater than 0.5 Mpc from each edge of the
  survey regions. This third condition ensures $>80$\% of the area of the
  1\,Mpc circle surrounding a galaxy is within the survey region.}

\end{enumerate}

We find that 7\,288 galaxies out of the 34\,102 galaxies
satisfy the second and third conditions, 
and overall we find 3\,536 galaxies to be isolated. 
As this study is focused on satellites hosted
by typical galaxies and not those in large groups and clusters, we
remove all 22 galaxies with $\log_{10} {\mathcal M}_{*} \ge 11.5$ that may be
brightest cluster galaxies, resulting in a sample of 3\,514 isolated
galaxies.

\subsection{Satellite Galaxies}   

After finding isolated systems we search for satellites around these
galaxies. We define satellite galaxies as those surrounding isolated
galaxies, which have a stellar mass that is at most one-third that of
the central galaxy $({\mathcal M}_{*} < 1/3{\mathcal M}_{*,cen})$, within a
projected radius of $R_{Proj} \le 500$\,kpc and $|c \Delta z| \le 500$ 
km\,s$^{-1}$. 
{\revone We chose a minimum host to satellite mass ratio of 3:1 as
  \cite{Hopkins2008b} find that a 3:1 mass ratio merger event is the
  limit for an $L^{*}$ disk to survive as a disk galaxy. A satellite
  with one-third the mass of its host will be $\sim 1.2$ mags fainter,
  assuming it has the same stellar mass-to-light ratio.}

Out of the 3\,514 isolated galaxies we find that 1\,426 host a total
of 2\,998 satellite galaxies. Noting that the more massive isolated
galaxies are more likely to have satellites, and the more massive
hosts are more likely to have multiple satellites, simply because of
the selection, we find that most (59.4 per cent) have no satellites
and the mean number of satellites per isolated galaxy is
0.85. Excluding the isolated galaxies which do not host satellites we
find that the mean number of satellites per central host is
2.10. Varying the projected radii and velocity differences in the
isolation and satellite search criteria results in slightly different
numbers of satellites per host and per isolated galaxy.

\subsection{Other Small Neighbours}

For the purposes of determining how the properties of satellite
galaxies depend on the host properties, we
produce a comparative sample of smaller neighbouring galaxies with
similar masses to the satellites, which satisfy the same criteria but
with projected distances in the range $0.5 \le R_{Proj} \le 1$ Mpc. To
ensure these neighbouring galaxies have no host, we check to see that
these neighbouring galaxies have no nearby galaxies which are greater
than 3 times the stellar mass of the small neighbour, within $R_{p} <
500$\,kpc and with $|c\Delta z| < 500$ km\,s$^{-1}$. As these galaxies are
selected in a similar way to the satellites, we consider this to be a
control sample. Surrounding the 3\,514 isolated galaxies we find a
total of 2\,304 of these small neighbours.

\section{Red and Blue Populations of Galaxies}

In this paper we make use of the well known colour bimodality of galaxies
\citep{Strateva2001, Baldry2004} to divide the hosts and satellite galaxies
into red and blue populations, with a colour-mass diagram (CMD). This enables
us to compare the properties between the populations, to examine how the
radial distribution of satellites depends on host colour, and to determine how
the red fraction of satellite galaxies varies as a function of both projected
radial distance from the host and stellar mass.

\subsection{Colour-Mass Distributions}

\begin{figure}
\centerline{
  \includegraphics[width=0.5\textwidth]{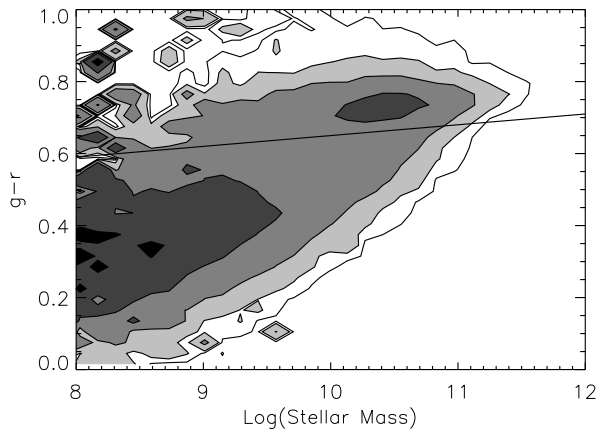}
}
\caption{{\it g-r} colour-mass diagram for the entire sample of
  34\,102 galaxies within $0.01 < z < 0.15$ from the {\it r}-band
  limited Main Survey. The data points are weighted by $1/V_{max}$ and
  represented as a logarithmic contour plot. Red and blue populations
  are separated by the line $(g-r) = 0.03 (\log_{10} {\mathcal M}_{*}) + 0.35$.}
\label{fig:CMD}
\end{figure}

Using $g-r$ calculated from the AUTO mags, we produce a CMD using the sample
of 34\,102 galaxies in the range $0.01 < z < 0.15$.  Figure~\ref{fig:CMD}
shows the CMD plotted as logarithmically spaced contours. Each data point is
weighted by $1/V_{max}$, where $V_{max}$ is the maximum comoving volume,
within which the galaxy could lie depending on its redshift and the limits of
the survey \citep{Schmidt1968}.
 
\begin{figure}
\includegraphics[width=0.5\textwidth]{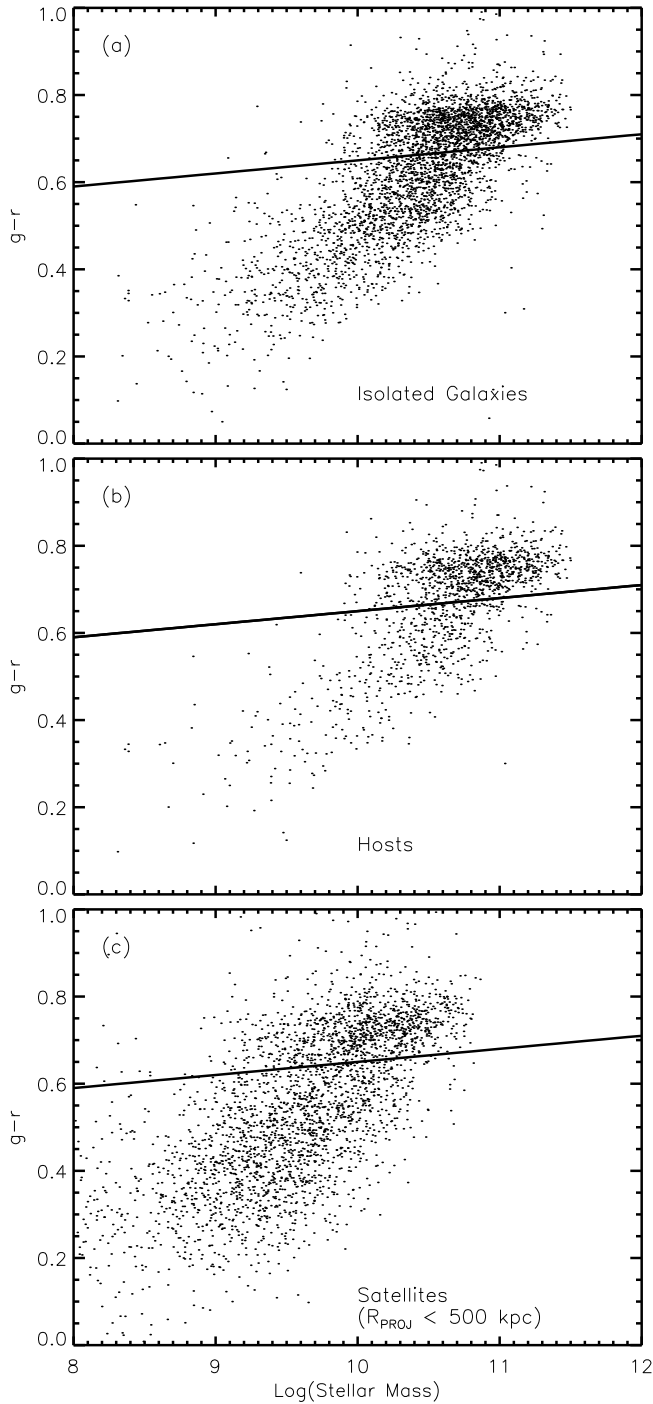}
\caption{{\it g-r} colour-mass diagrams for {\bf (a)} the 3\,514 isolated
  galaxies {\bf (b)} the 1\,426 central hosts and {\bf (c)} 2\,998
  satellites. Red and blue populations are separated by the line
  $(g-r) = 0.03(\log_{10} {\mathcal M}_{*}) + 0.35$.}
\label{fig:colour-mass}
\end{figure}

Colour bimodality can clearly be seen in Figure~\ref{fig:CMD}, and we choose
to separate the red and blue populations using a straight line with the
equation:
\begin{equation}
   (g-r) = 0.03 (\log_{10} {\mathcal M}_{*}) + 0.35.
\end{equation} 
Figure~\ref{fig:CMD} looks almost identical to the CMD in figure~A1 of
\cite{VDBosch2008}, produced using stellar masses derived from
\cite{Bell2003}. CMDs for the isolated galaxies, central hosts and satellites
with projected distances of $R_{Proj} \le 500$ kpc can be seen in
Figure~\ref{fig:colour-mass}, which clearly shows that the majority of
isolated galaxies which host satellites are red (906 out of 1\,426 hosts). Out
of the 2\,998 satellites we find 1\,097 red and 1\,901 blue galaxies. By
further dividing the sample into red/blue satellite/host systems, we find that
41.2 per cent (1\,235) of the satellites are blue with red hosts, 33.3 per
cent (998) are red satellites with red hosts, 22.2 per cent (666) are blue
satellites with blue hosts, and only 3.3 per cent (99) are red satellites with
blue hosts. Overall the blue fraction of satellites is 63.4 per cent, which
rises to 87.1 per cent for blue hosts compared to 55.3 per cent for red hosts.
 
\subsection{Luminosity and Mass Distributions}

\begin{figure*}
\includegraphics[width=0.42\textwidth]{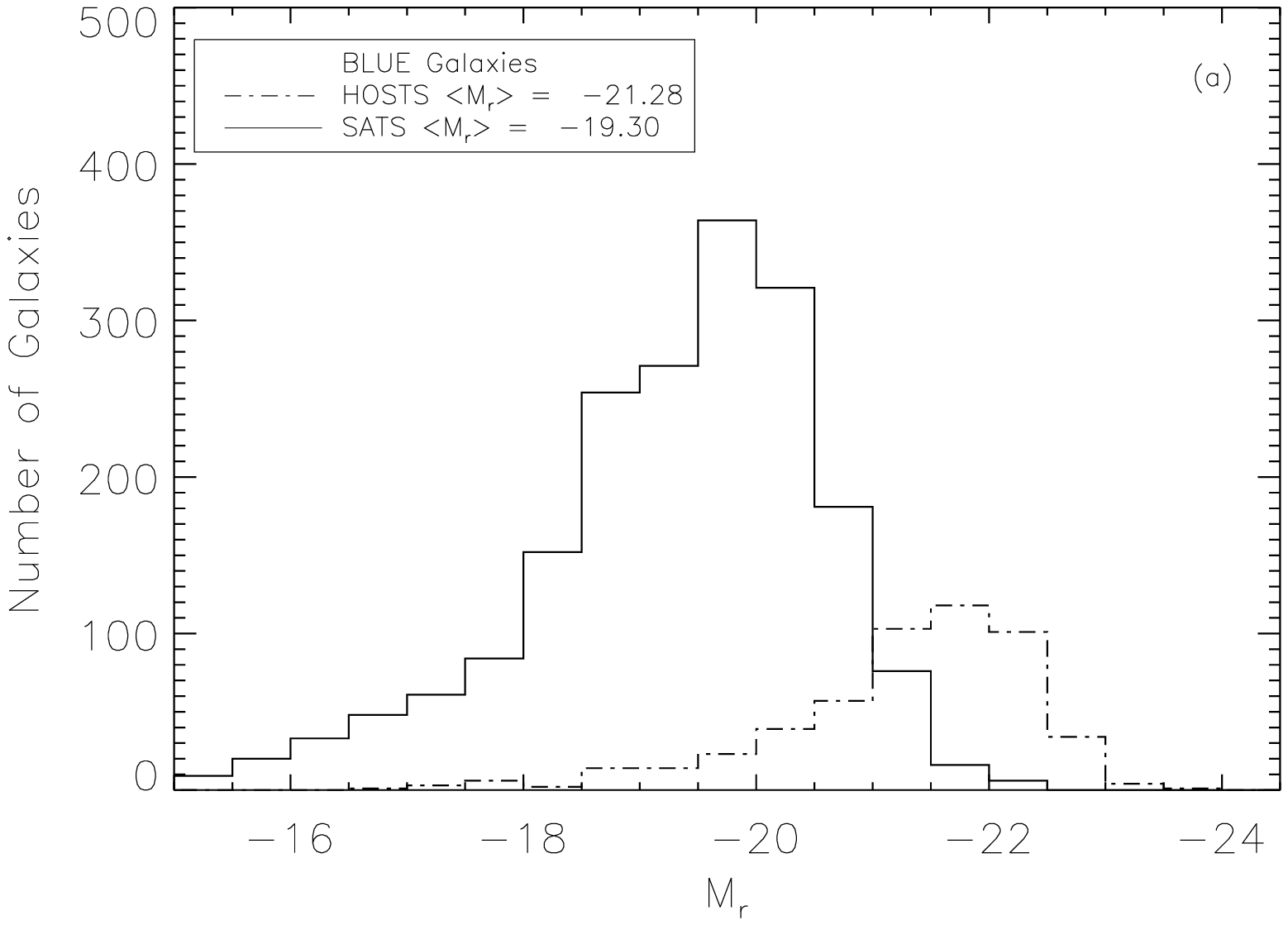}
\includegraphics[width=0.42\textwidth]{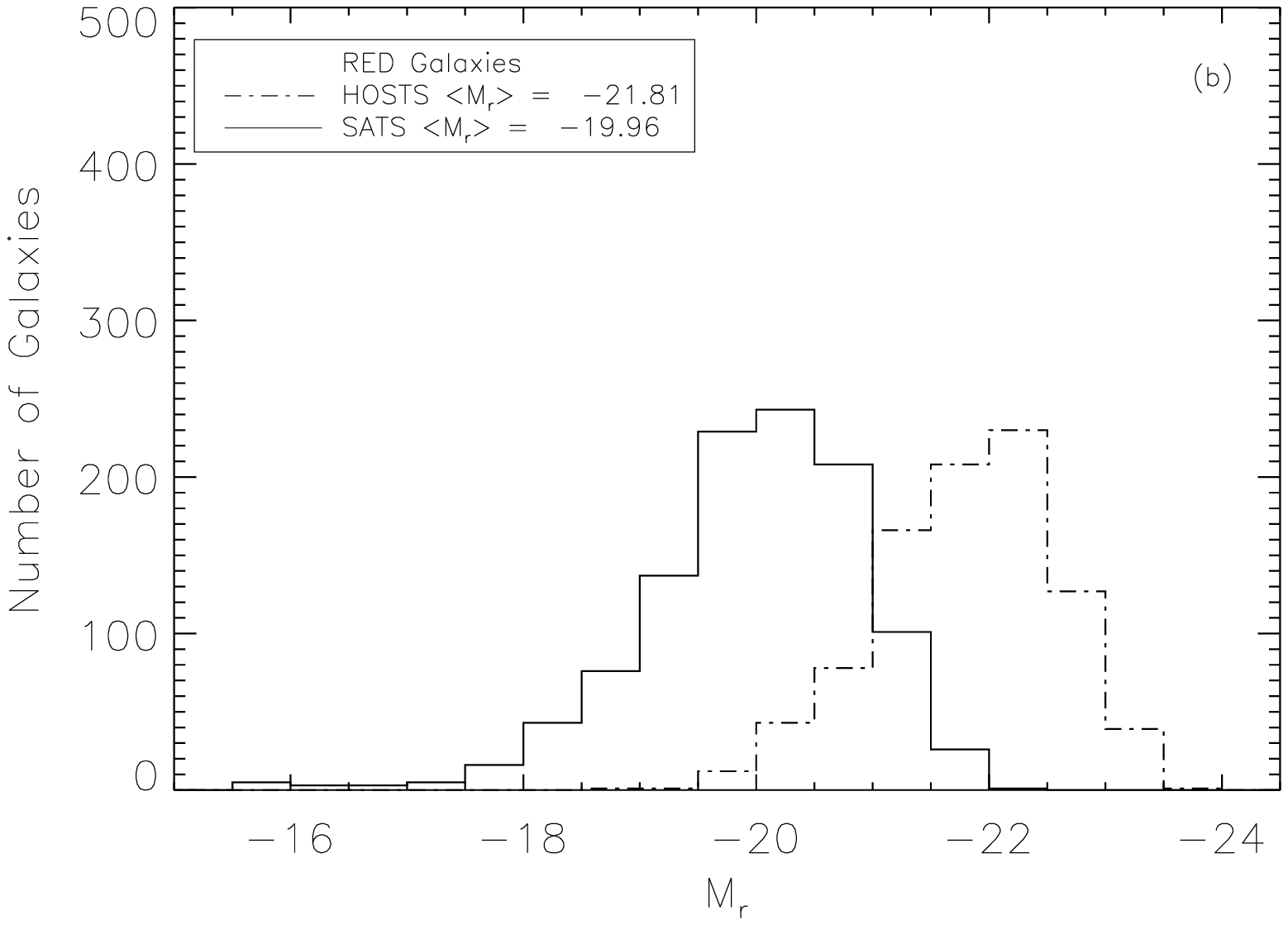}
\includegraphics[width=0.42\textwidth]{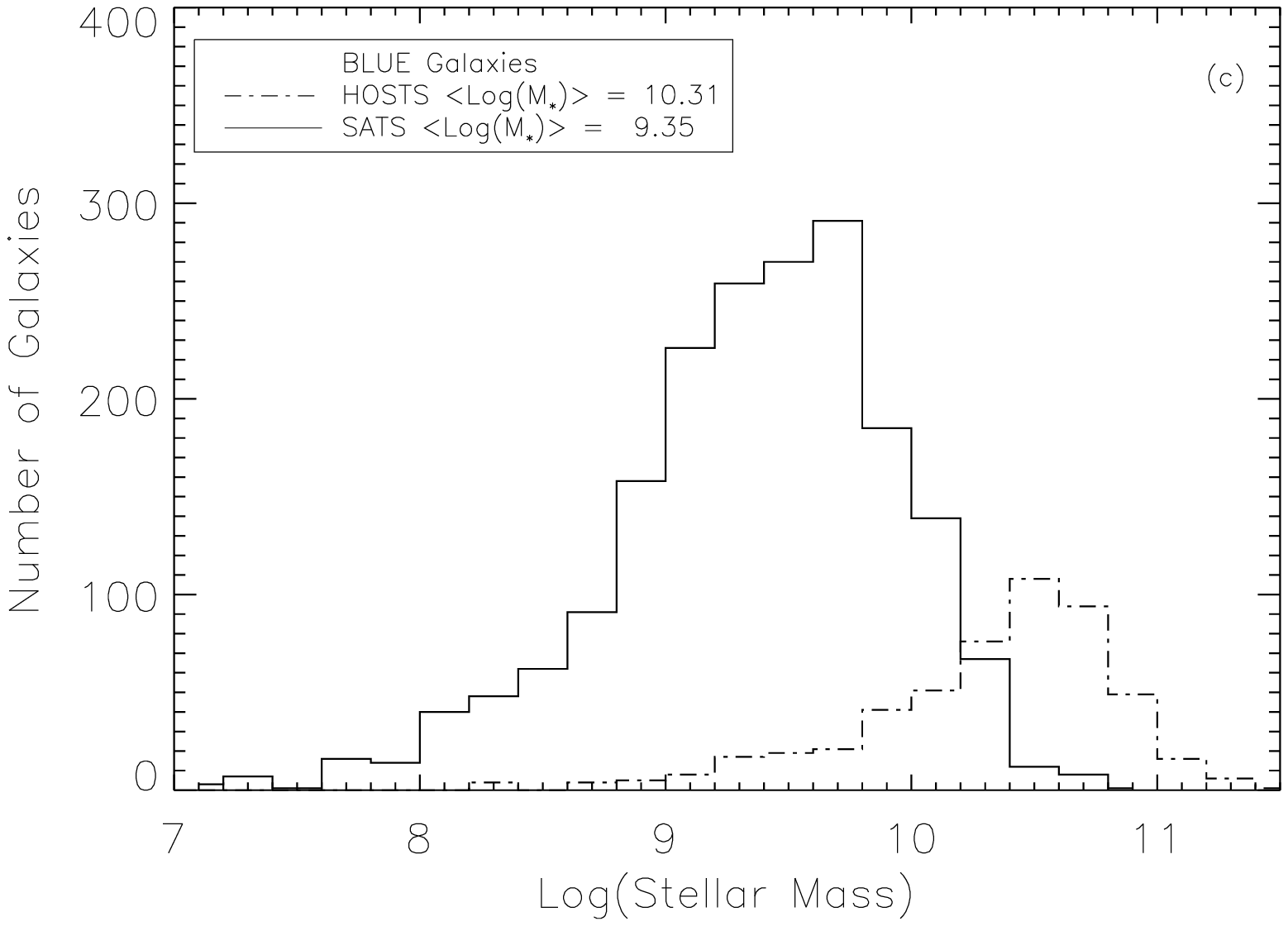}
\includegraphics[width=0.42\textwidth]{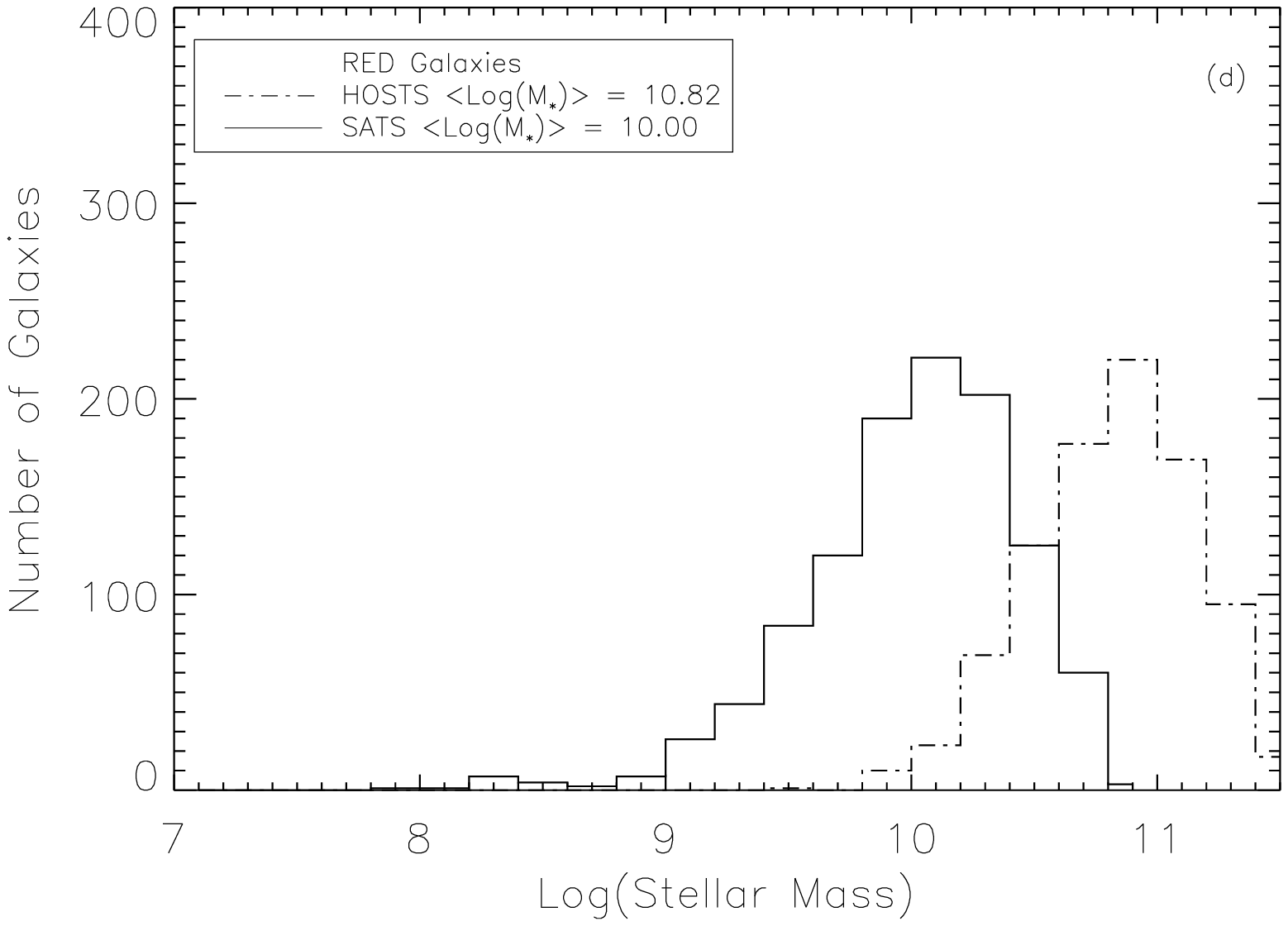}
\caption{Histograms showing the distribution of {\it r}-band absolute
  magnitudes {\bf (a,b)} and stellar masses {\bf (c,d)} for the blue 
  and red populations
  of satellites with $R_{Proj} \le 500$ kpc (solid lines) and host galaxies
  (dashed lines).}
\label{fig:rdist}
\includegraphics[width=0.42\textwidth]{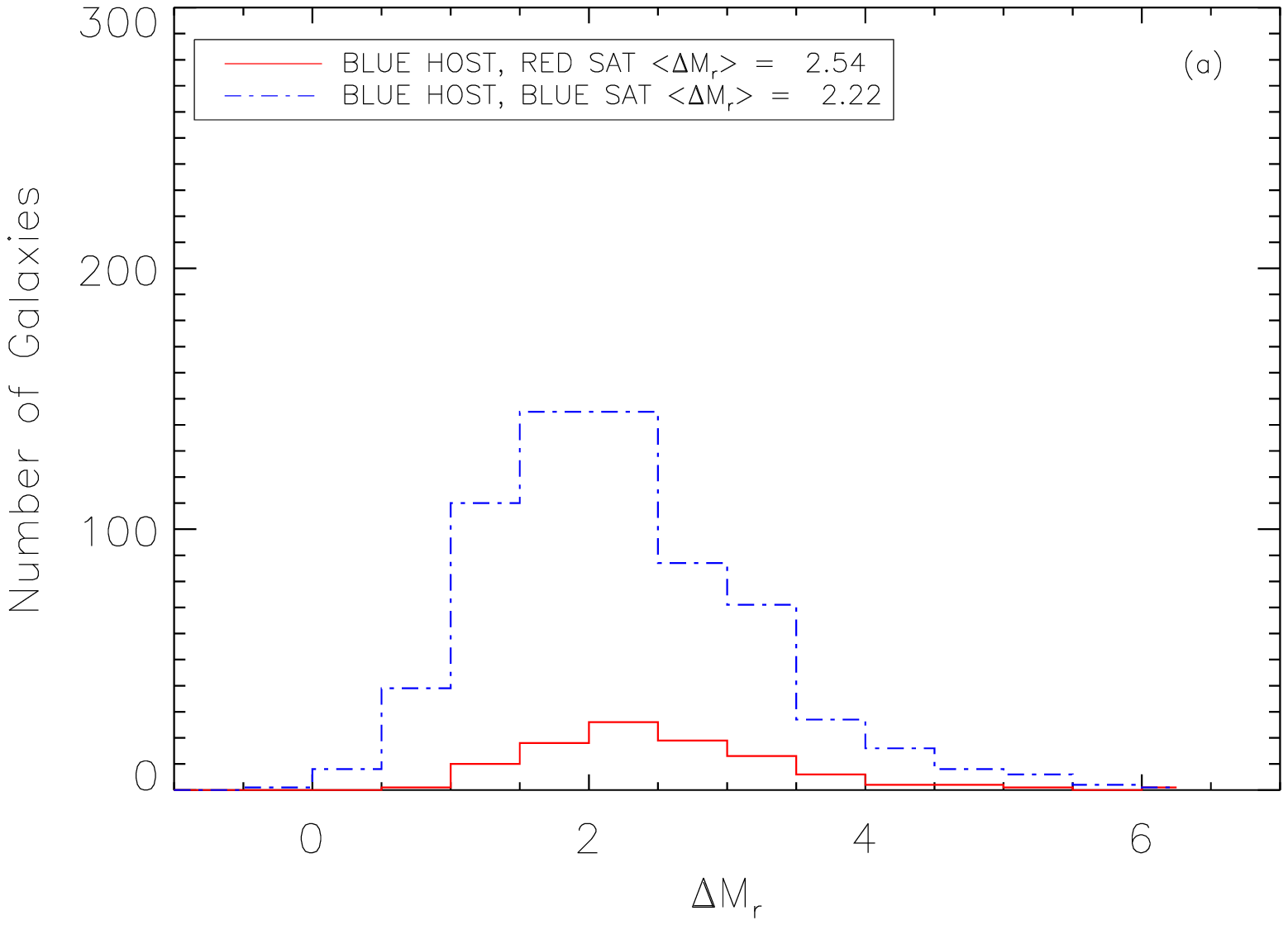}
\includegraphics[width=0.42\textwidth]{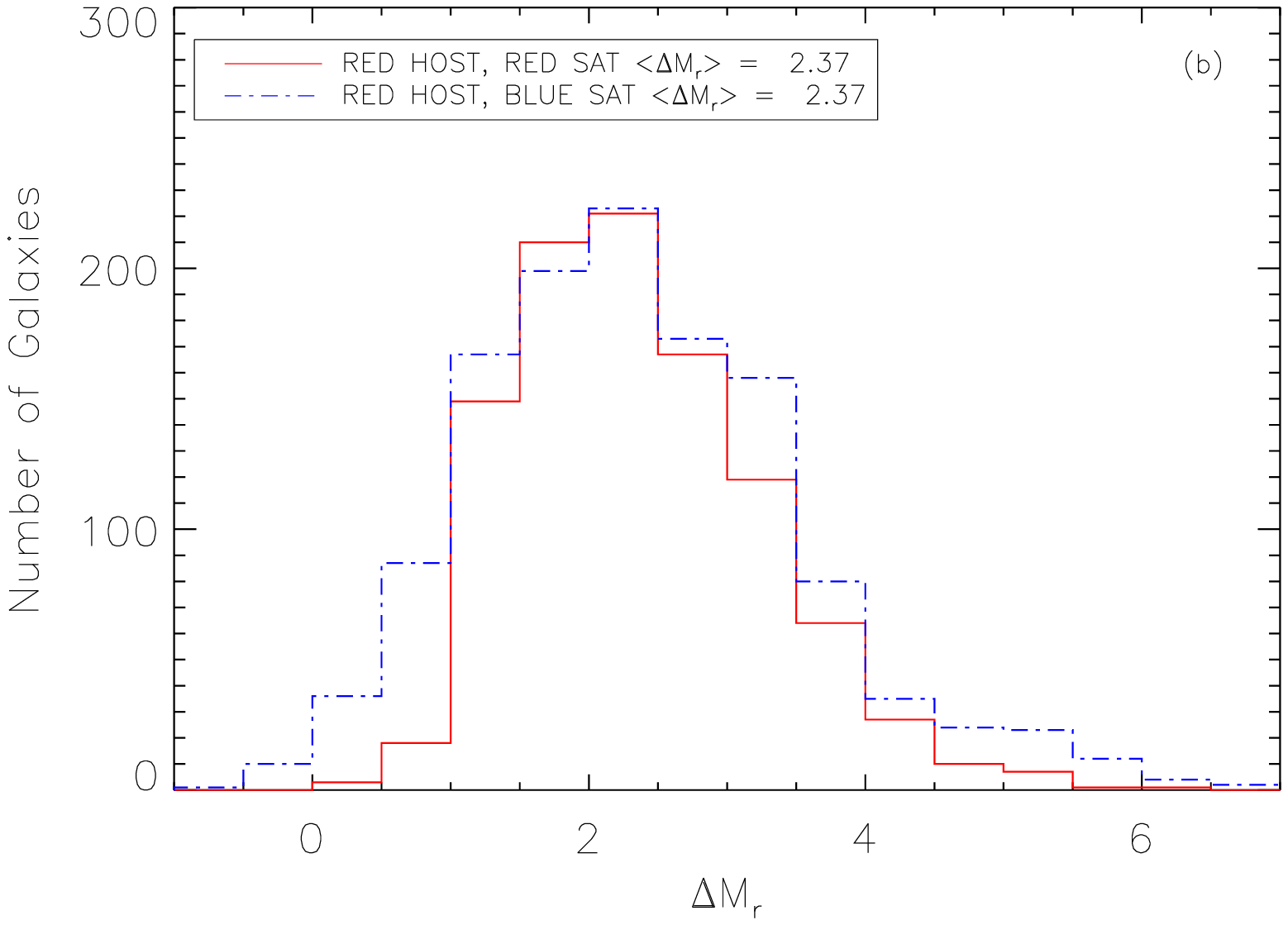}
\includegraphics[width=0.42\textwidth]{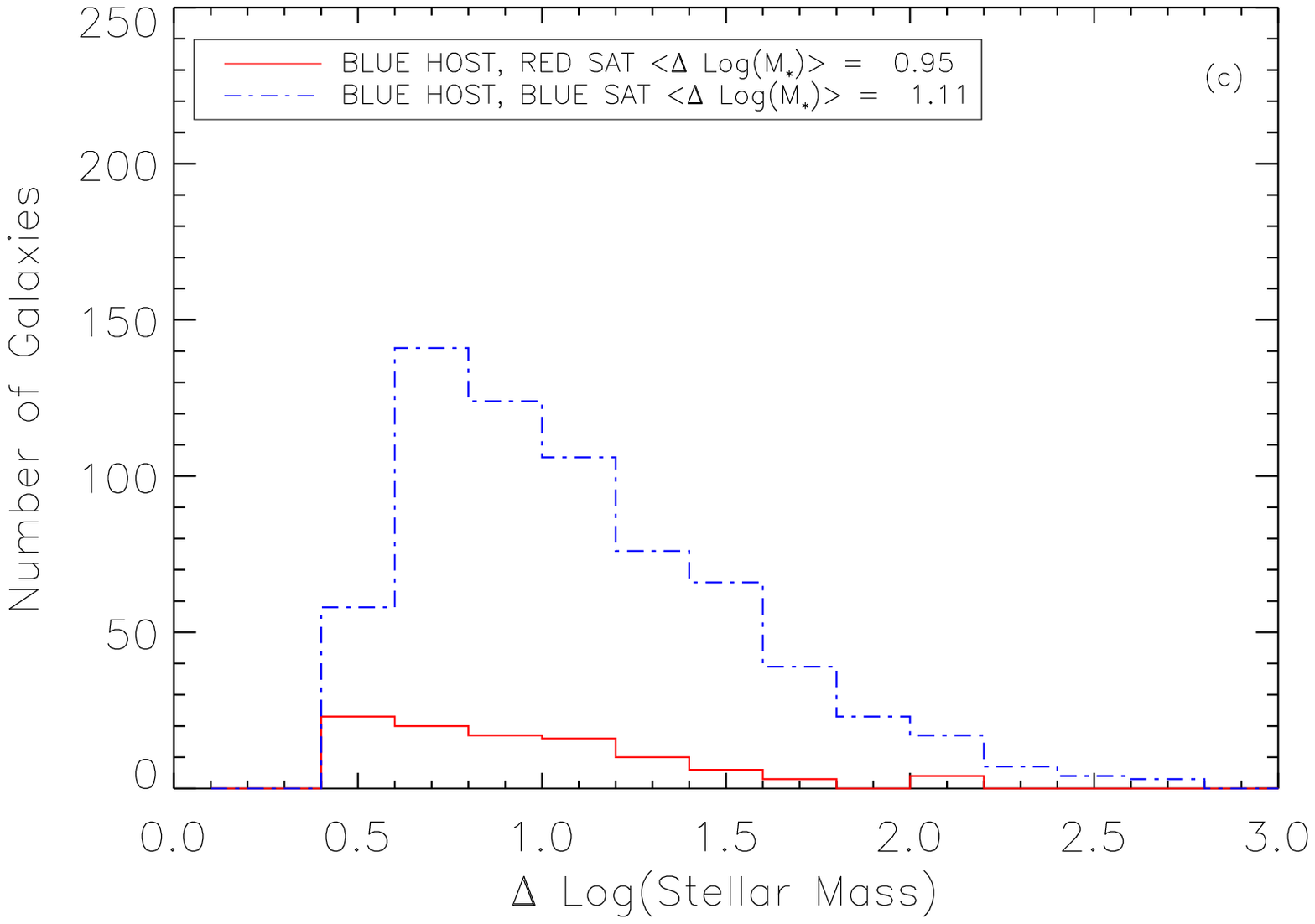}
\includegraphics[width=0.42\textwidth]{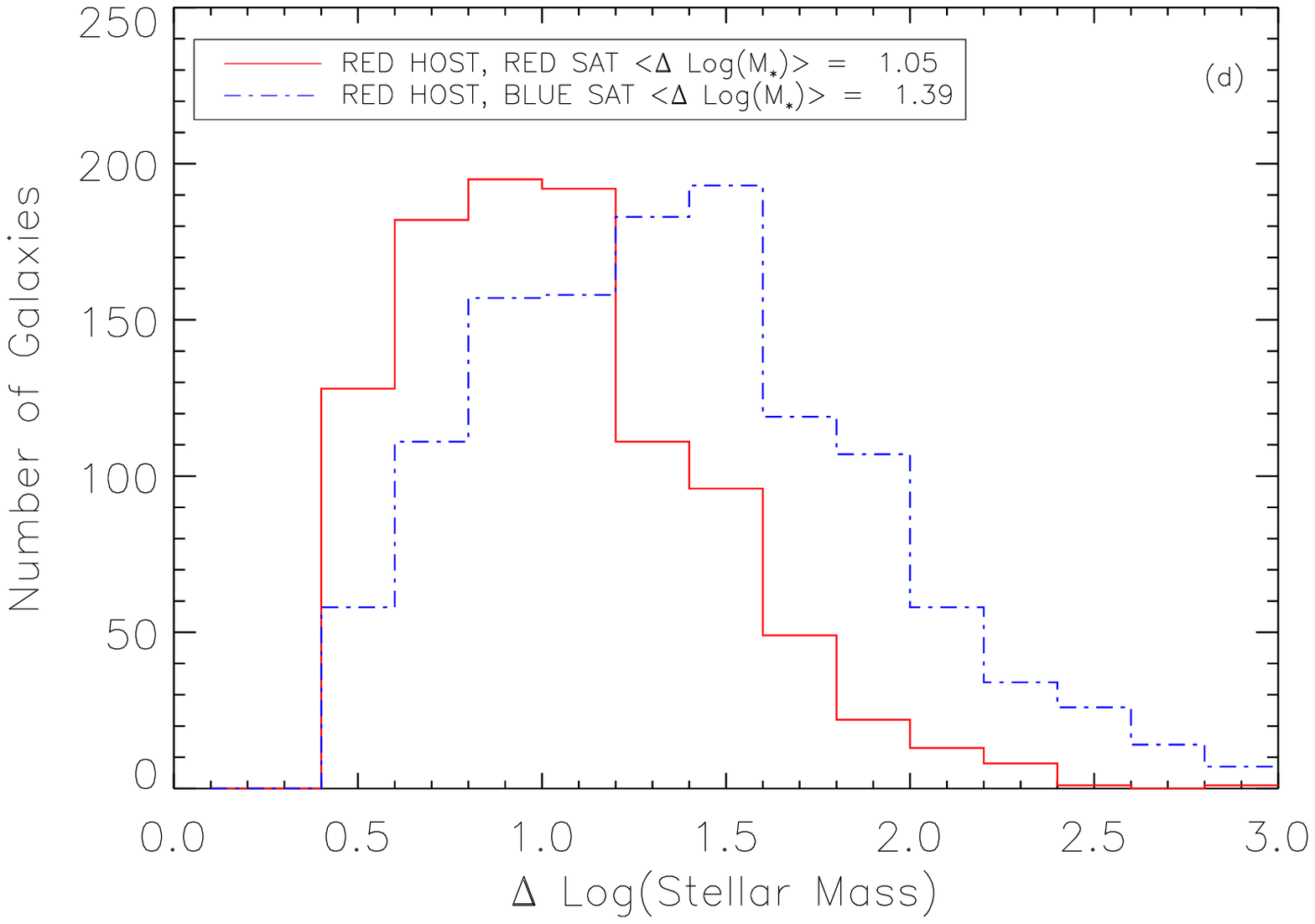}
\caption{Distributions of the differences in the absolute {\it r}-band
  magnitude {\bf (a,b)} and stellar masses {\bf (c,d)} between the red 
  and blue satellites
  ($R_{Proj} \le 500$ kpc) and their red and blue central hosts. Histograms
  for red satellites are shown as dotted lines, blue satellites are shown as
  dotted-dashed lines.}
\label{fig:rdist2}
\end{figure*}

In this section we compare the properties of the different samples of
galaxies. In Figure~\ref{fig:rdist} we show the distributions in luminosity
and stellar mass for red and blue hosts and satellites. This shows, as
expected, the red population on average being more luminous and massive than
the blue population. We calculate mean stellar masses and absolute {\it
  r}-band magnitudes for red host galaxies as $\log_{10} {\mathcal M}_{*} =
10.81$ and $M_{r} = -21.81$ respectively, compared to $\log_{10} {\mathcal
  M}_{*} = 10.31$ and $M_{r} = -21.28$ for the blue hosts. The latter is
similar to the value for the Milky Way, and the Schechter break ($M^{*}$) of
blue galaxies (Loveday et al.\ 2011 in preparation). Similarly for the red
satellites we calculate means of $\log_{10} {\mathcal M}_{*} = 10.00$ and
$M_{r} = -19.96$, and $\log_{10} {\mathcal M}_{*} = 9.35$ and $ M_{r} =
-19.30$ for the blue. Both populations are typically more massive and luminous
than the Large Magellanic Cloud.

Comparing the host galaxies to isolated galaxies without satellites,
we find that both the blue and red galaxies with satellites are more
massive and more luminous than those without, which is most likely a
result of our satellite selection as the more massive isolated
galaxies have a larger chance of hosting a satellite. The average mass
and luminosity of the red galaxies without satellites is $\log_{10}
{\mathcal M}_{*} = 10.61$ and $M_{r} = -21.39$ compared to the blue
galaxies which have $\log_{10} {\mathcal M}_{*} = 10.12$ and $M_{r} =
-20.94$.

The mean masses and luminosities of the small neighbours are very
similar to those of the satellites and we find means of $\log_{10}
{\mathcal M}_{*} = 10.77$ and $M_{r} = -20.17$ for red galaxies and
$\log_{10} {\mathcal M}_{*} = 9.42$ and $M_{r} = -19.48$ for the
blue. This is reassuring as it means that the small neighbours do
indeed have similar properties to the satellites and can be considered
a `control' sample, allowing comparisons between the two.

Other quantities useful for comparison with cosmological models
include the stellar mass and luminosity differences between hosts and
satellites. In Figure~\ref{fig:rdist2} we show histograms of these
quantities for the blue and red hosts.

Due to our satellites having stellar masses that are less than 1/3 of their
host mass, $\Delta \log_{10} {\mathcal M}_{*}$ has a minimum value of 0.477,
and we find a maximum of $\Delta \log_{10} {\mathcal M}_{*} =3$ (1/1000th the
mass of the host). We find find that the average satellite in this study is
$\sim$ 1/10th of the mass of the host and calculate mean logarithmic mass
differences of 1.05 and 1.39 for red and blue satellites with red hosts, and
0.95 and 1.11 for red and blue satellites with blue hosts.

As for luminosity difference, we find that satellites are typically
$\sim 2$ magnitudes fainter than their hosts. Red and blue satellites
of red hosts, are both found to have mean magnitude differences of
2.37, and red and blue satellites with blue hosts have magnitude
differences of 2.54 and 2.21. We also find a small fraction of
satellites which have a negative value of $\Delta M_{r}$ and are
brighter than their hosts, which would be missed by other studies.
{\revone In other words, when considering red hosts, they can be more
  than three times more massive than their blue satellites yet be 
  of similar luminosity.}

\clearpage

\section{Projected Density of Satellites}

To investigate the radial distribution of the satellites we
calculate the projected density per host given by:
\begin{equation}
\Sigma(R) = \frac{N_{Sat}(R)}{\pi N_{Host}({{R_{2}}^{2}-{R_{1}}^{2}})}
\end{equation}
where $N_{Sat}(R)$ is the number of satellites found within a shell
bounded by inner and outer projected radii, $R_{1}$ and $R_{2}$,
from the host and $N_{Host}$ is the number of host galaxies and $R$ is
the midpoint of the bin.

\begin{figure}
\includegraphics[width=0.47\textwidth]{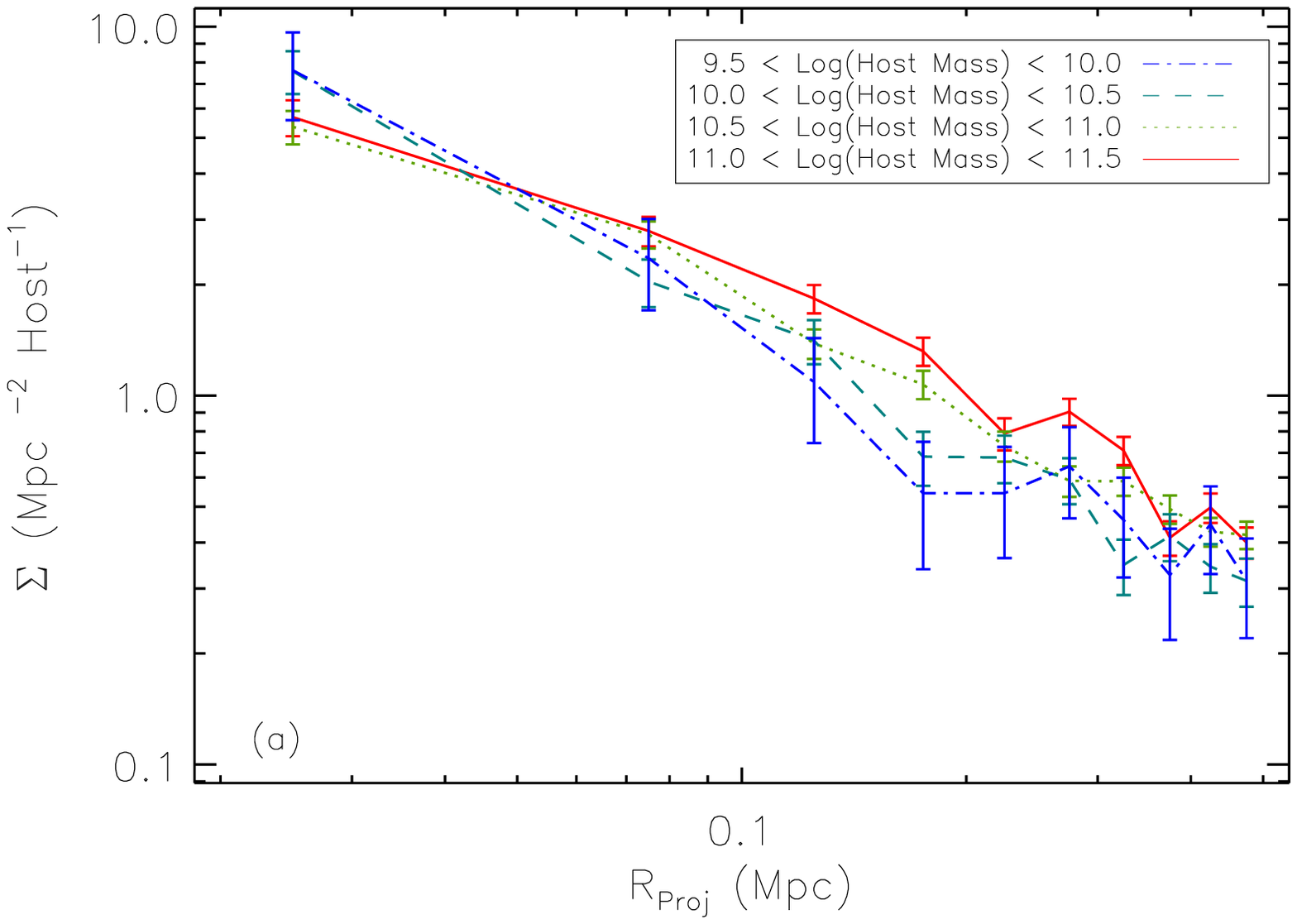}
\includegraphics[width=0.47\textwidth]{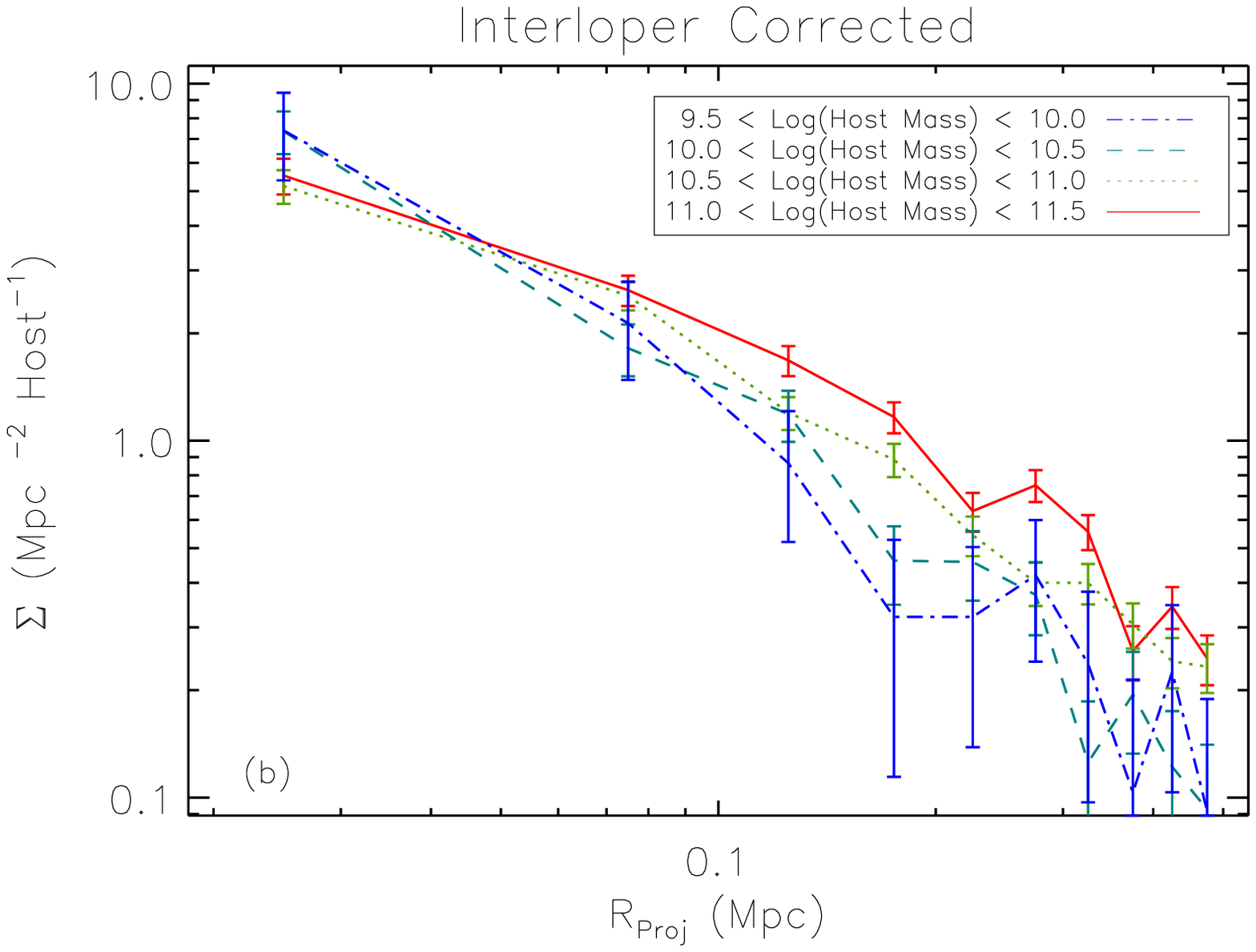}
\caption{{\bf (a)}: Projected densities of satellites with $R_{Proj}
  \le 500$ kpc, divided into four host stellar mass bins, using
  projected radius bins of width $\Delta R = 0.05$ Mpc. {\bf (b)}: The
  same as above corrected for interlopers. The error bars shown are
  Poissonian.}
\label{fig:rprojmass}
\end{figure}

Using projected radius bins of width $\Delta R = 0.05$ Mpc and
dividing the satellites into four host stellar mass bins; $9.5 <
\log_{10} {\mathcal M}_{*} < 10.0$, $10.0 < \log_{10} {\mathcal M}_{*}
< 10.5$, $10.5 < \log_{10} {\mathcal M}_{*} < 11.0$ and $11.0 <
\log_{10} {\mathcal M}_{*} < 11.5$, we produce
Figure~\ref{fig:rprojmass}(a) which shows how the projected density
of galaxies varies as a function of projected distance. We fit the
projected density with a power law, parametrized by a slope, $\alpha$,
and normalisation $A$:
\begin{equation}
\Sigma(R) =  A (R/{\rm Mpc})^\alpha .
\label{eqn:ar}
\end{equation}
with $A$ in Mpc$^{-2}$ as $\Sigma$ is in Mpc$^{-2}$. 
     
\begin{table*}
\caption{Best fitting A and $\alpha$ parameters (Equation~\ref{eqn:ar}) for
  the projected densities in Figure~\ref{fig:rprojmass} and
  Figure~\ref{fig:rprojcol}, corrected and uncorrected for
  interlopers. {\revtwo The uncertainties quoted are derived from
    least-squares straight-line fits to $\log \Sigma$ versus $\log R$ with
    the data points shown in the figures.}}
\label{tab:projfits}
\begin{center}
\begin{tabular}{lccccc} \hline
Host Stellar Mass Range &$N_{\rm Sat\,tot}$       & A$_{\rm{un\,corr}}$ (Mpc$^{-2}$)& $\alpha_{\rm{un\,corr}}$ & A$_{\rm{Int\,corr}}$ (Mpc$^{-2}$)& $\alpha_{\rm{Int\,corr}}$\\
\hline 
$9.5 \le \log_{10} {\mathcal M}_{*} \le 10.0$   &111 &$0.14 \pm 0.03$ & $-1.06 \pm 0.10$ & $0.05 \pm 0.01$  & $-1.40 \pm 0.10$\\
$10.0 \le \log_{10} {\mathcal M}_{*} \le 10.5$  &457 &$0.13 \pm 0.01$ & $-1.09 \pm 0.05$ & $0.04 \pm 0.04$  & $-1.46 \pm 0.05$\\
$10.5 \le \log_{10} {\mathcal M}_{*} \le 11.0$  &1\,225 &$0.20 \pm 0.01$ & $-0.94 \pm 0.03$ & $0.10 \pm 0.01$  & $-1.14 \pm 0.04$\\       
$11.0 \le \log_{10} {\mathcal M}_{*} \le 11.5$  &1\,140 &$0.23 \pm 0.02$ & $-0.94 \pm 0.04$ & $ 0.14 \pm 0.01$ & $-1.09 \pm 0.04$\\  
\hline   
ALL BLUE HOSTS  &700 &$0.14 \pm 0.01$ & $-1.04 \pm 0.04$& $0.05 \pm 0.01$& $-1.38 \pm 0.04$\\  
ALL RED HOSTS   &2\,223 &$0.21 \pm 0.01$ & $-0.94 \pm 0.03$& $0.12 \pm 0.01$& $-1.13 \pm 0.03$\\  
\hline
\end{tabular}
\end{center}
\end{table*} 

\begin{figure}
\includegraphics[width=0.47\textwidth]{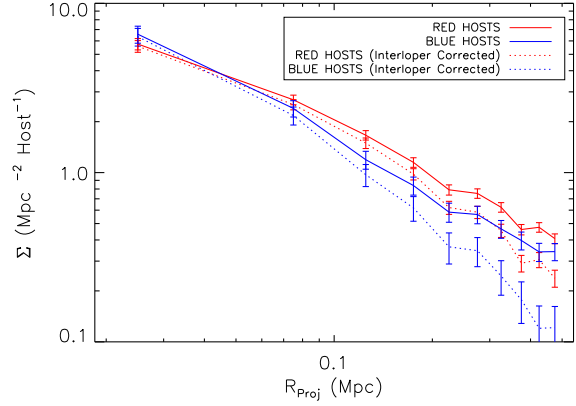}
\caption{Projected densities of satellites with $R_{Proj} \le 500$
  kpc, for the red and blue hosts with masses in the range $9.5 \le
  \log_{10} {\mathcal M}_{*} \le 11.5$ , using projected radius bins
  of width $\Delta R = 0.05$ Mpc. The interloper corrected values are
  connected by dotted lines. The error bars are Poissonian.}
\label{fig:rprojcol}
\end{figure}

In Table~\ref{tab:projfits} we show the best-fitting parameters and errors
for each of the curves, and find that there is little variation in the
best-fitting slope, which steepens slightly from $\alpha = -0.94 \pm
0.04$ for satellites with hosts in the mass range $11.0 < \log_{10}
{\mathcal M}_{*} < 11.5$, through to $\alpha = -1.06 \pm 0.03$ for
satellites with hosts in the mass range $9.5 < \log_{10} {\mathcal M}_{*}
< 10.0$. Dividing the satellites into those with red and blue hosts
(Figure~\ref{fig:rprojcol}), with host masses between $ 9.5 < \log_{10}
{\mathcal M}_{*} < 11.5$, we find a best fitting slope of $\alpha = -0.94
\pm 0.03$ for satellites with red hosts compared to those with blue
host which have $\alpha = -1.04 \pm 0.04$. 
{\revtwo These slopes suggest} that the blue, low mass host galaxies have
slightly more centrally concentrated satellites. The slopes however are
affected by interlopers, galaxies which are mistaken as satellites by
projection and not actually bound to their hosts.

In order to correct for interlopers we use a simple method of
subtracting the average projected density of the small neighbours,
with projected radii in the range $0.5 < R_{Proj} < 1.0$ Mpc, from the
projected density of the satellites in each bin:
\begin{equation}
\Sigma(R)_{{\rm Int\,Corr}} = \Sigma(R)_{{\rm Sat}} 
  - \langle\Sigma(0.5 < R < 1.0)_{{\rm small\,neigh}}\rangle \: .
\end{equation}
{\revtwo Note that this assumes that these galaxies in the 0.5--1\,Mpc annuli
  represent the average large-scale or `super cluster' environment around the
  hosts, and the correction will be an overestimate in hosts where a
  significant bound satellite population extends beyond 0.5\,Mpc, in
  particular for our most massive host sample (11.0--11.5).}

The result of applying this interloper correction and fitting the
slope for each of the host mass ranges (and colour) can also be seen
in Table~\ref{tab:projfits}, which shows that the best-fitting slope now
significantly varies with host mass and steepens from $\alpha = -1.09
\pm 0.04$ for satellites with hosts in the mass range $11.0 <
\log_{10} {\mathcal M}_{*} < 11.5$ through to $\alpha = -1.4 \pm 0.1$
for satellites with hosts in the mass range $9.5 < \log_{10} {\mathcal
  M} < 10.0$. For the red and blue hosts, we obtain interloper
corrected slopes of $\alpha = -1.13 \pm 0.03$ and $\alpha = -1.38 \pm
0.04$ respectively. The results of this inteloper subtraction can be
seen in Figure~\ref{fig:rprojmass}(b) and as the dotted lines
in Figure~\ref{fig:rprojcol}.

An observed decline in the projected density, 
which overall can be approximated by $\Sigma(R) \propto R^{-1}$, is similar
to other studies that investigate the radial distribution of
satellites. For example, \cite{Lorrimer1994} using the Centre for
Astrophysics redshift survey, find a slope of $\alpha = -0.91 \pm
0.05$ for satellites with magnitudes in the range $-16 < B < -18$,
after correcting for background galaxies, which is slightly shallower
than the interloper corrected slopes we obtain. Furthermore, they
divide their satellite sample into early- and late-type hosts, to find
that satellites with early-type hosts are more centrally concentrated
than late types, finding slopes of $\alpha = -1.01 \pm 0.1$ for
satellites of early-type hosts and $\alpha = -0.78 \pm 0.05$ and for
the late-type hosts, which is the opposite to what we find. Although
our slope for satellites of red hosts is consistent with their result
for early-type hosts, our result for the satellites of blue galaxies 
differs from their findings for late-type hosts.

More recently, \cite{Chen2006} find an interloper corrected slope of
$\alpha = -1.58 \pm 0.11$, using a galaxy sample from SDSS DR4. Here
satellites are defined as galaxies with $M_{r} < -18.0$, which are 2
mags fainter than their hosts, with $R_{Proj} \le 500$ kpc and $\Delta
V \le 500$ km\,s$^{-1}$ and isolated galaxies are defined as those with
$M_{r} \le -20.0$, have no nearby galaxies within $\Delta m \le 2$
mag, $\Delta V \le 1000$ km\,s$^{-1}$ and $R_{Proj} \le 500$ kpc. Using
these same selection criteria and a larger data sample from SDSS DR6,
\cite{Chen2008} find that the interloper corrected slopes of
satellites of red and blue hosts are almost identical and obtain
values of $\alpha = -1.54 \pm 0.12$ and $\alpha = -1.50 \pm 0.12$ for
satellites of blue hosts and red hosts, respectively. Although these
values are similar to our slopes for the blue and low-mass hosts, they
are quite different for the red and high-mass hosts.  {\revone
  \cite{Chen2008} and others have used $r$-band luminosity for
  isolation criteria. For red hosts, this is significantly more
  stringent than using an estimated stellar mass that allows for
  equally luminous blue satellites. This could explain some of these
  differences.}

\cite{Sales2007} using a sample of galaxies from the semi-analytic
Millennium Simulation find a slope of $\alpha = -1.55 \pm 0.08$, for
satellites again that are at least 2 magnitudes fainter than their
hosts, within $R_{Proj} \le 500$ kpc and $\Delta V \pm 500$ km\,s$^-1$.

Investigating the satellites of 34 elliptical galaxies
\cite{Madore2004} find a decline with a slope of $\alpha = -0.5$ out
to $R_{Proj} \simeq 150$ kpc, which flattens out at greater projected
radii. Although the slope they obtain is again inconsistent with our results,
this may be due to the small numbers of galaxies involved. They do
note that there is a flattening in the projected density at $R_{Proj}
\ge 300$ kpc, which can also be seen in our data.

Finally, \cite{Ann2008} using SDSS DR5 data, define satellites as
galaxies with $M_{r} < -18.0$, which are more than 1 mag fainter in
the {\it r}-band than their hosts, within $R_{Proj} \le 1$ Mpc and
$\Delta V \le 1000$ km\,s$^{-1}$. They also use an isolation criterion, so
that isolated galaxies are required to have $M_{r} < -19.0$, and no
neighbours within $\Delta m \le 1$ mag of the host, within a $R_{Proj}
\le 500$ kpc and $\Delta V \pm 500$ km\,s$^{-1}$, producing a sample of
2\,254 hosts with 4\,986 satellites. Without correcting for
interlopers they fit slopes for satellites within $R_{Proj} = 200$ kpc
of their hosts, and obtain values of $\alpha \sim -1.8$ for both the
late- and early-type satellites of late-type hosts and for the
early-type satellites of early-type hosts. For the late-type satellites of
early-type hosts a shallower slope of $\alpha \sim -1.5$ is found.

{\revtwo The shallower slopes for the high-mass hosts compared to low-mass
  hosts measured over the range 0.05--0.5\,Mpc are expected if satellite
  density profiles are related to dark-matter profiles (\citealt*{NFW96},
  hereafter NFW; surface density profile given by \citealt{Bartelmann1996}).
  The higher-mass hosts would be expected to have larger NFW scale radii and
  therefore shallower dark-matter profiles over the radial range probed by our
  study.} 

\section{Red Fraction of Satellite Galaxies}

After measuring the projected density of satellites, we
determine how the red fraction of satellites varies as a function of
mass and projected distance from the hosts.
{\revone It should be noted that the most robust results, 
free of significant selection effects, are the differences between
the satellites and small neighbours.}

\subsection{Red Fraction as a Function of Projected Distance}

\begin{figure}
\includegraphics[width=0.47\textwidth]{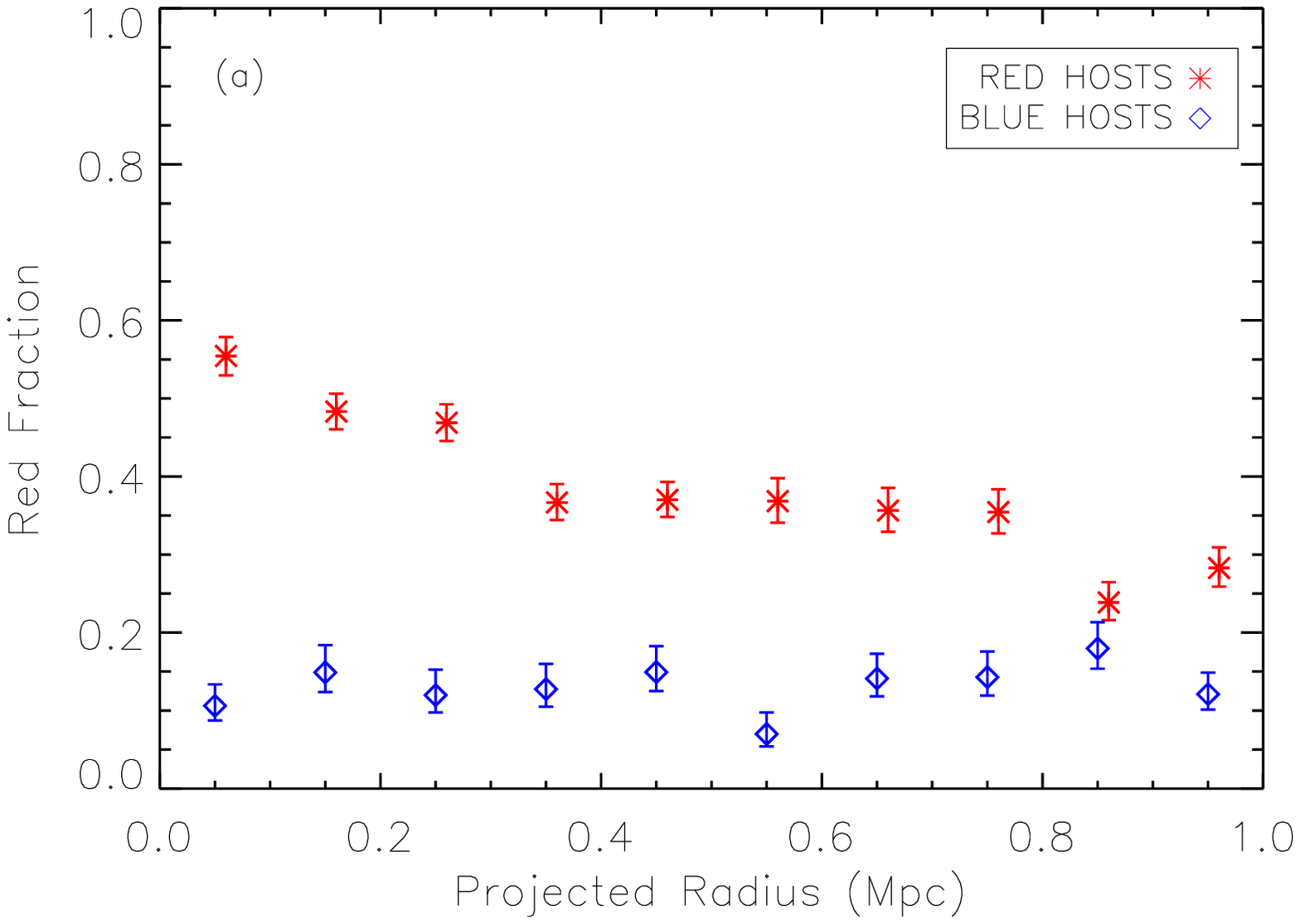}
\includegraphics[width=0.47\textwidth]{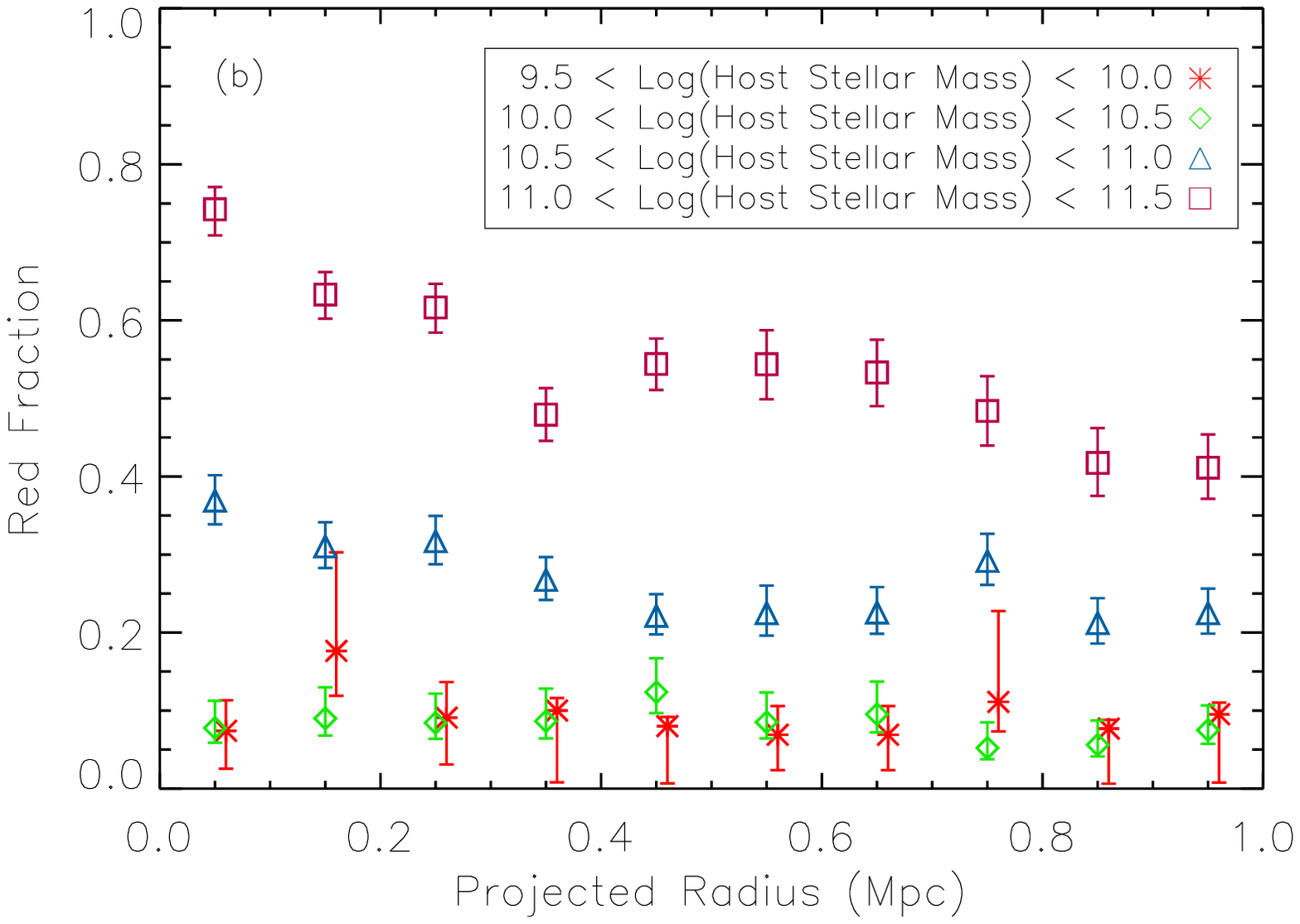}
\caption{{\bf (a)}: The red fraction of satellite galaxies as a
  function of projected radius for the red and blue hosts. {\bf (b)}:
  The red fraction of satellite galaxies as a function of projected
  radius, for four different host stellar mass ranges. We define
  galaxies with $R_{Proj} \le 0.5$ Mpc as satellites and galaxies with
  $0.5 \le R_{Proj} \le 1.0$ Mpc as small neighbouring galaxies which
  have been selected in a similar way. The error bars shown are
  1$\sigma$ beta distribution confidence intervals.}
\label{fig:rfhost}
\end{figure}

 To measure the red fraction of satellites ($R_{Proj} \le 0.5$ Mpc)
 and small neighbours ($0.5 \le R_{Proj} \le 1 $ Mpc) we divide the
 number counts of red galaxies by the total number of galaxies in bins
 of projected radius of width $\Delta R_{Proj} = 0.1$ Mpc.  {\revone By
   dividing the sample of satellites into those with red and blue
   hosts we produce Figure~\ref{fig:rfhost}(a). By
 dividing the sample of satellites into four bins of host stellar mass
 as before, we produce Figure~\ref{fig:rfhost}(b). The error bars on the red
 fraction of galaxies in all plots are 1$\sigma$ beta distribution
 confidence intervals explained in detail in
 \cite{Cameron2010}. These are an improvement over other methods for
 estimating binomial confidence intervals, such as Poisson errors, which
 are liable to misrepresent the degree of statistical uncertainty for
 small samples, or fractions that approach values of 0 or 1.}

{\revone From Figure~\ref{fig:rfhost}(a) it is clear that the red fraction of
  satellites of red hosts is significantly higher than those with blue
  hosts. The average red fraction of all the points between $0.05 \le
  R_{Proj} \le 1$ Mpc for the satellites of red hosts is 0.38 compared
  to 0.13 for the blue hosts. The red fraction of satellites for red
  hosts increases by a factor of 1.49 from 0.37 at $R_{Proj} = 0.45$
  Mpc to 0.55 at $0.05$ Mpc. For the satellites of blue hosts no trend
  in the red fraction as a function of radius is seen.}

Similar trends are reflected in Figure~\ref{fig:rfhost}(b) as expected. 
The red
fraction of satellites and small neighbours increases significantly as
a function of host mass, from an average red fraction of all the
points between $0.05 \le R_{Proj} \le 1$ Mpc of 0.09 for galaxies
associated with hosts with $9.5 \le \log_{10} {\mathcal M}_{*} \le
10.0$, 0.08 for galaxies of hosts with $10.0 \le \log_{10} {\mathcal
  M}_{*} \le 10.5$, 0.27 for galaxies of hosts with $10.5 \le
\log_{10} {\mathcal M}_{*} \le 11.0$, through to 0.54 for galaxies of
hosts with $11.0 \le \log_{10} {\mathcal M}_{*} \le 11.5$. The highest
two host mass ranges show that an upturn in the red fraction occurs
within $R_{Proj} \sim 500$ kpc. For the highest host mass bin $11.0
\le \log_{10} {\mathcal M} \le 11.5$, the red fraction of satellites
increases by a factor of 1.4 from 0.54 at $R_{Proj} = 0.45$ Mpc to
0.74 at $0.05$ Mpc and for satellites with host masses of $10.5 \le
\log_{10} {\mathcal M}_{*} \le 11.0$, with the red fraction increasing
by a factor of 1.7 from 0.22 at $R_{Proj} = 0.45$ Mpc to 0.37 at
$0.05$ Mpc. No trend in the red fraction as a function of radius is
observed for the 2 lowest host mass bins.

\begin{figure*}
\includegraphics[width=0.47\textwidth]{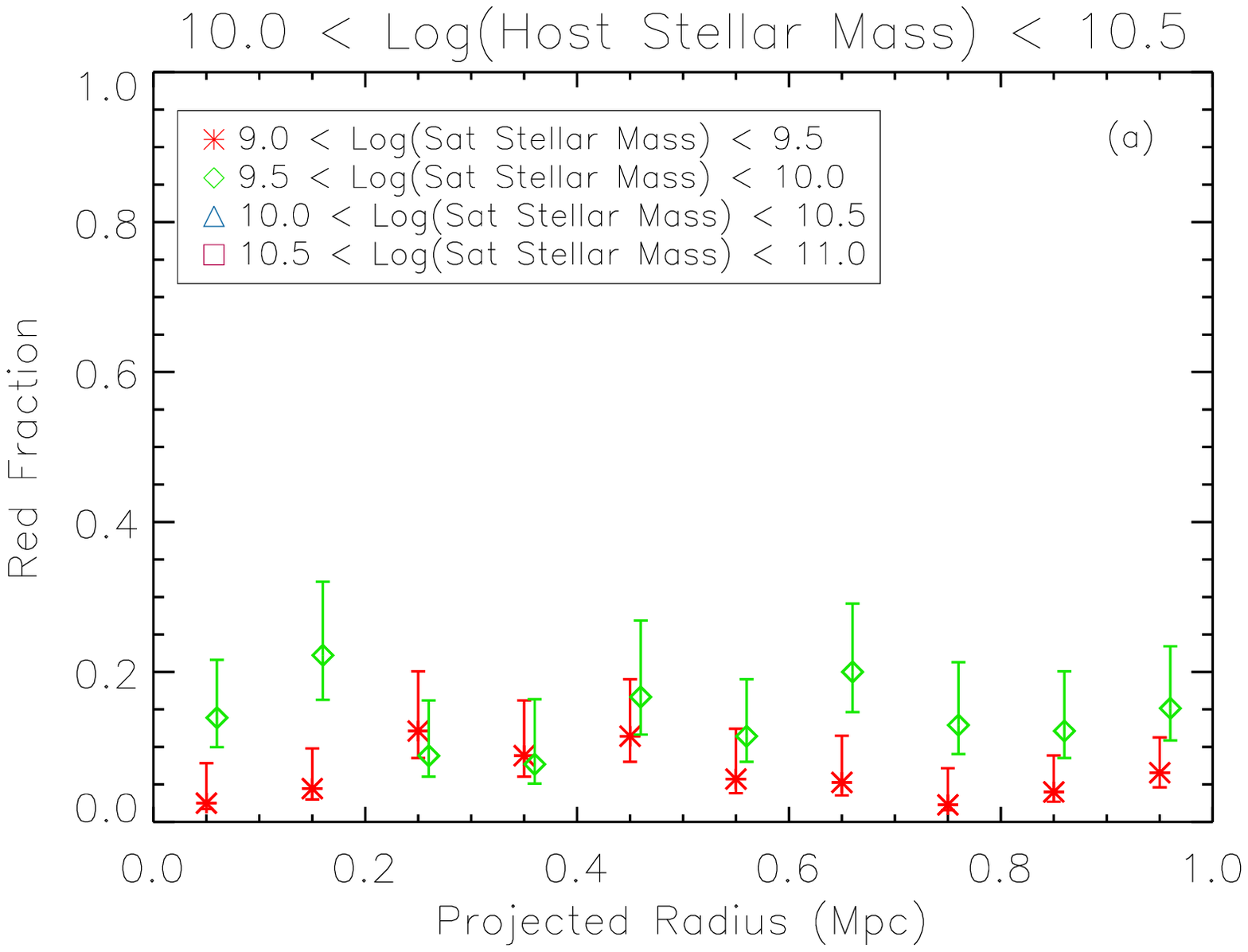}
\includegraphics[width=0.47\textwidth]{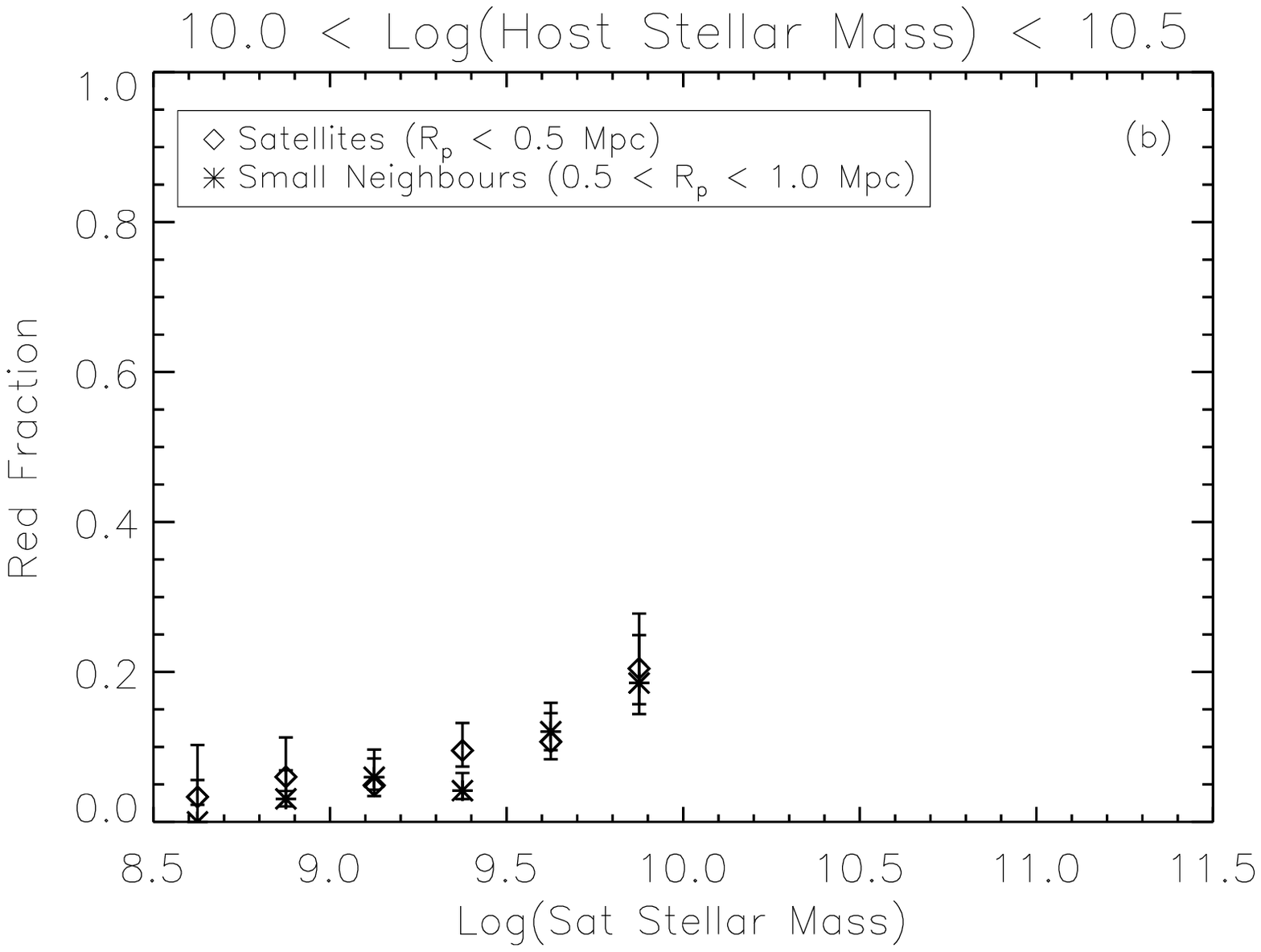}
\includegraphics[width=0.47\textwidth]{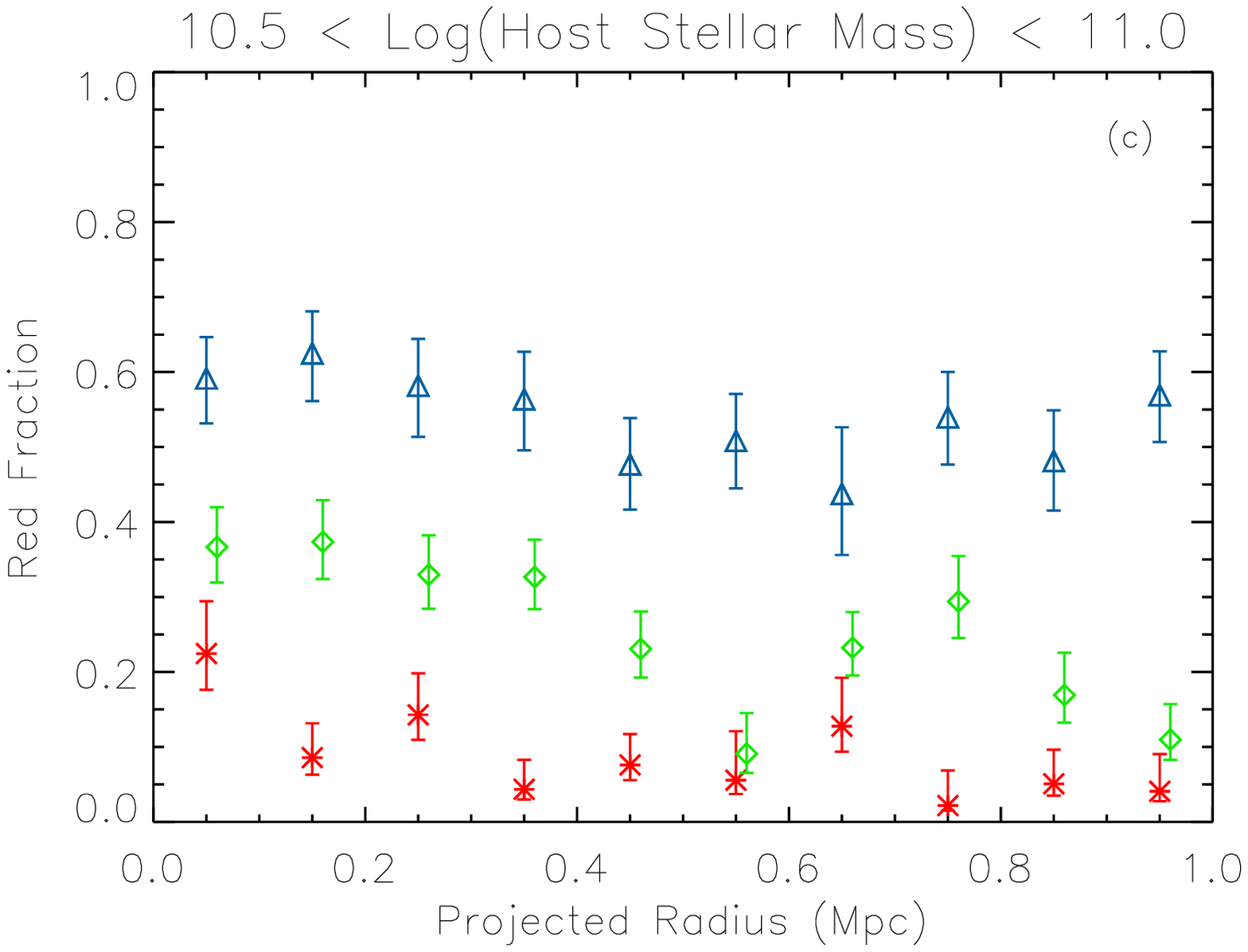}
\includegraphics[width=0.47\textwidth]{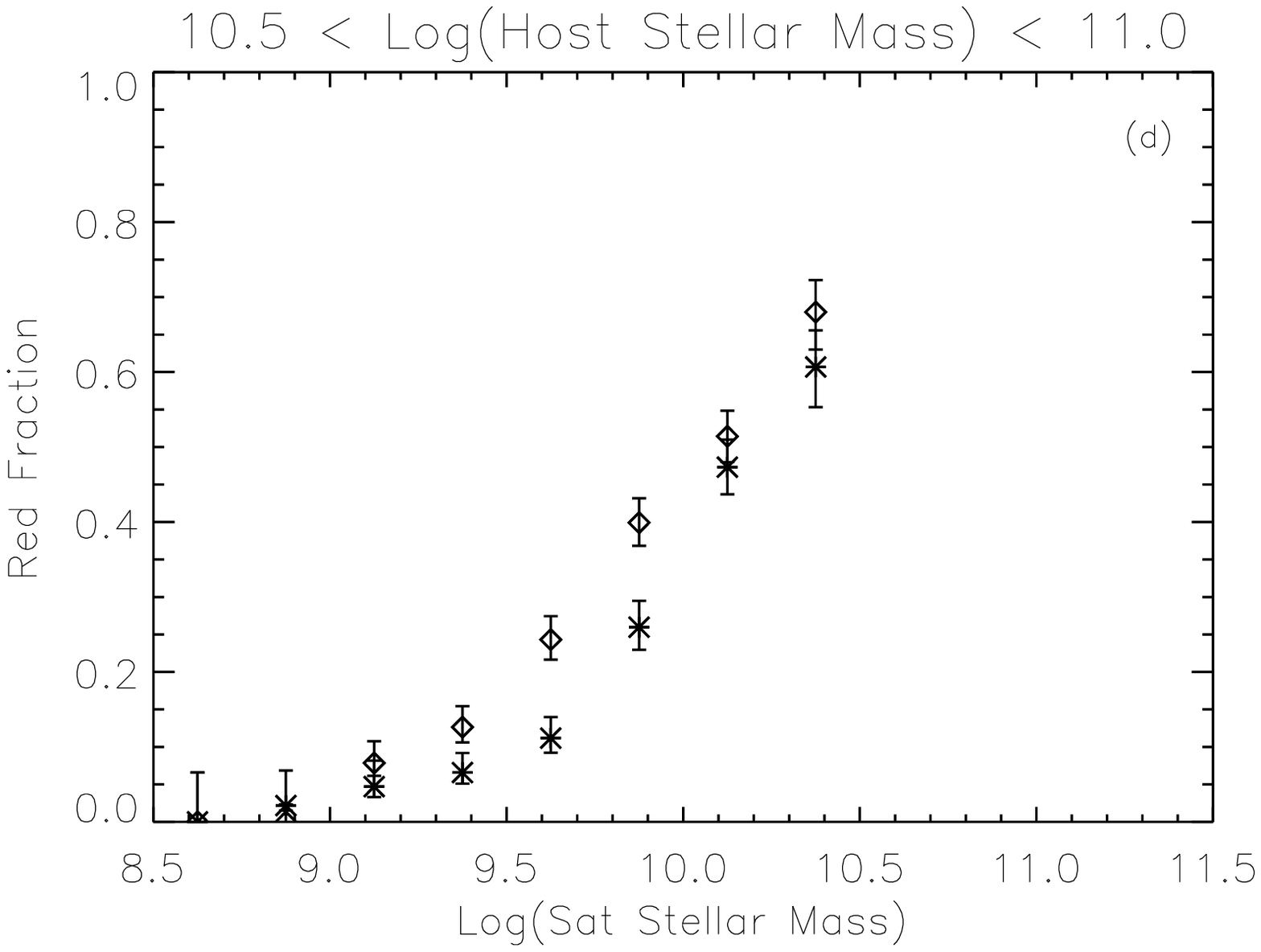}
\includegraphics[width=0.47\textwidth]{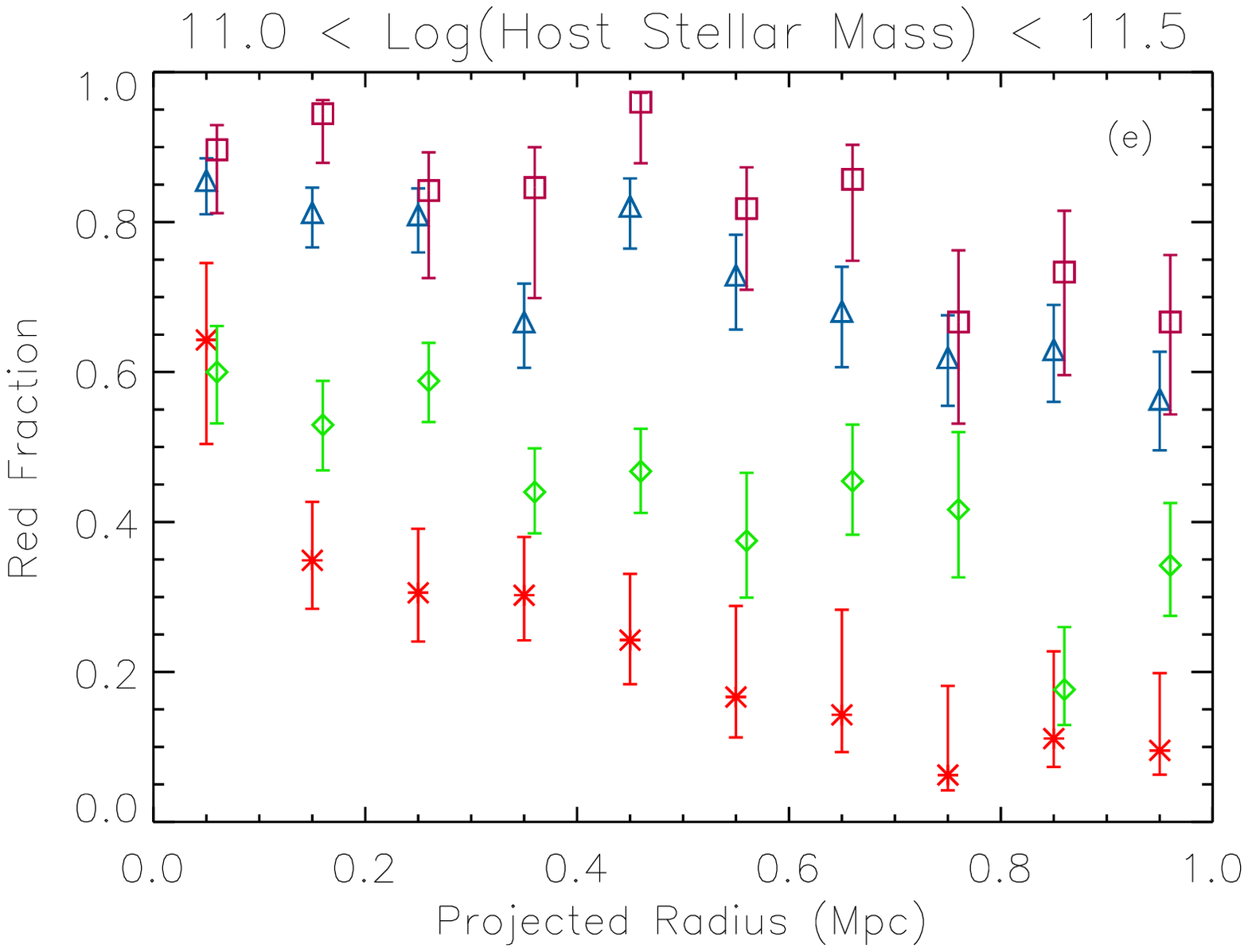}
\includegraphics[width=0.47\textwidth]{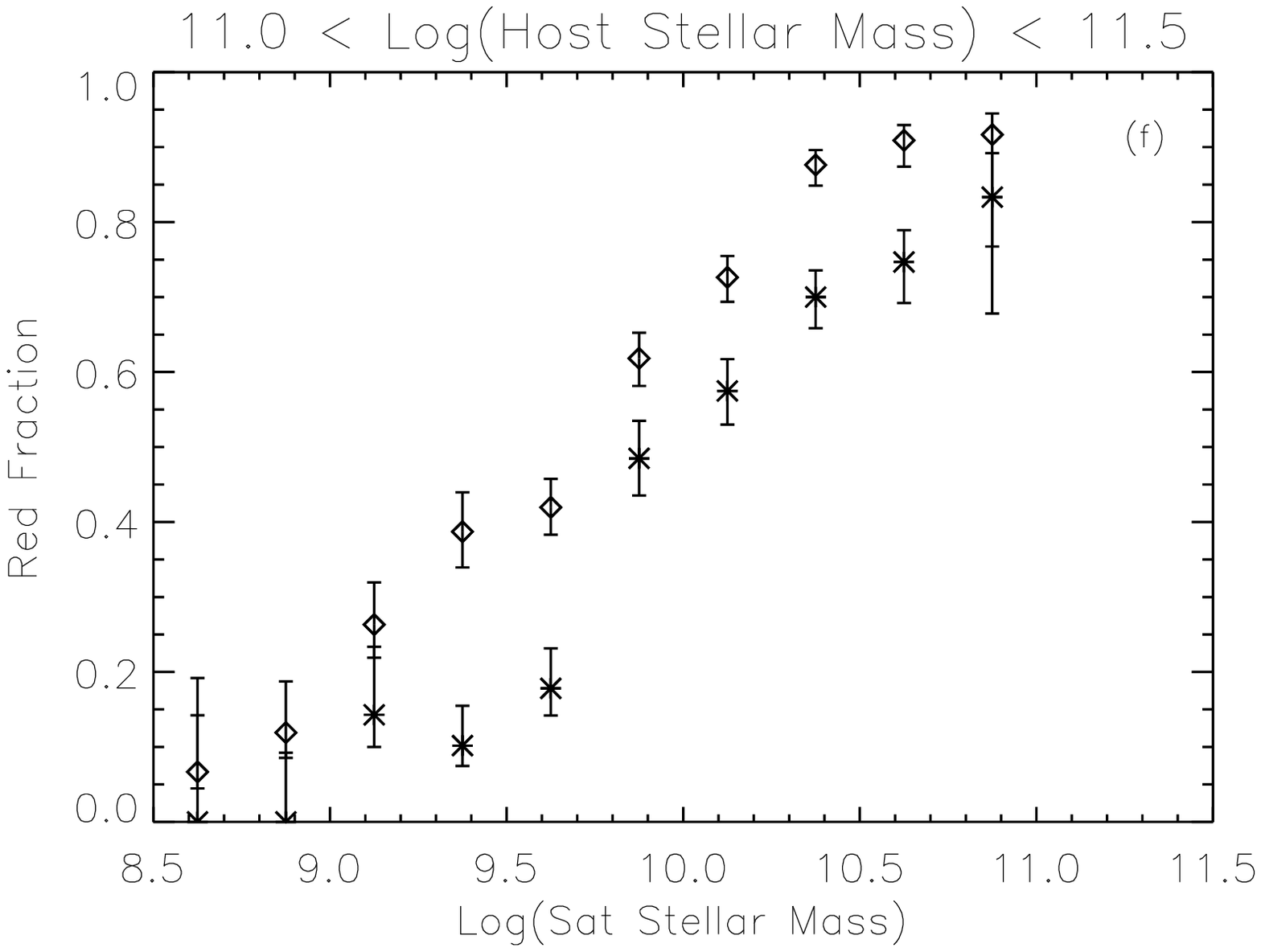}
\caption{Left panels {\bf (a,c,e)} show the red fraction of satellite galaxies
  as a function of projected radius and satellite stellar mass, shown for 3
  different central host stellar mass ranges. Host masses of $10.0 \le
  \log_{10} {\mathcal M}_{*} \le 10.5$, $10.5 \le \log_{10} {\mathcal M}_{*}
  \le 11.0$ and $11.0 \le \log_{10} {\mathcal M}_{*} \le 11.5$ with satellites
  of stellar masses $9.0 \le \log_{10} {\mathcal M}_{*} \le 9.5$ (red stars),
  $9.5 \le \log_{10} {\mathcal M}_{*} \le 10.0$ (green diamonds), $10.0 \le
  \log_{10} {\mathcal M}_{*} \le 10.5$ (blue triangles) and $10.5 \le
  \log_{10} {\mathcal M}_{*} \le 11.0$ (purple squares).  Right panels {\bf
    (b,d,f)} show the red fraction of satellites with $R_{Proj} \le 0.5$ Mpc
  and small neighbours with $0.5 \le R_{Proj} \le 1.0$ Mpc as a function of
  satellite mass, for hosts in the same mass ranges opposite. The errors shown
  are 1$\sigma$ beta distribution confidence intervals.}
\label{fig:rfmain}
\end{figure*}

Following on from this we investigate how the red fraction varies as a
function of satellite mass, by subdividing galaxies in the host mass bins
above further into satellite mass bins. {\revtwo It is important to do this in
  order to verify that the satellite trends are not simply due to mass
  segregation, and to separate `environment quenching' from `mass quenching'
  \citep{Peng2010}.}  Figure~\ref{fig:rfmain}(a,c,e) shows how the red
fraction varies as a function of radius, with the data separated into 3
different plots for three different host mass bins, and then subdivided into
satellite mass bins of $9.5 \le \log_{10} {\mathcal M}_{*} \le 10.0$, $10.0
\le \log_{10} {\mathcal M}_{*} \le 10.5$, $10.5 \le \log_{10} {\mathcal M}_{*}
\le 11.0$ and $11.0 \le \log_{10} {\mathcal M}_{*} \le 11.5$, depending on the
host mass bin. {\revone The mean numbers of satellites and small neighbours in
  each bin are 36.2, 65.6 and 48.5 for those with host masses in the ranges
  $10.0 \le \log_{10} {\mathcal M}_{*} \le 10.5$, $10.5 \le \log_{10}
  {\mathcal M}_{*} \le 11.0$ and $11.0 \le \log_{10} {\mathcal M}_{*} \le
  11.5$ respectively. } In Table~\ref{tab:averagered} we show the average red
fractions of satellites and small neighbours for each of the different host
and satellite mass ranges. From these it can be seen that:

\begin{table*}
\caption{Average red fraction of satellites and small neighbours of
  galaxies in Figure~\ref{fig:rfmain}. The error quoted is the
  standard error.}
\label{tab:averagered}
\begin{center}
\begin{tabular}{lcccc} \hline
Host Mass Range     &Satellite Mass Range     &  Mean Satellite Red Fraction & Mean Small Neighbour Red Fraction\\
\hline 
$10.0 \le \log_{10} {\mathcal M}_{*} \le 10.5$       &$9.0 \le \log_{10} {\mathcal M}_{*} \le 9.5$       &$0.08 \pm 0.02$ &$0.04 \pm 0.01 $\\
& $9.5 \le \log_{10} {\mathcal M}_{*} \le 10.0$                                                     &$0.14 \pm 0.03$ &$0.14 \pm 0.02 $\\
\hline
$10.5 \le \log_{10} {\mathcal M}_{*} \le 11.0$       & $9.0 \le \log_{10} {\mathcal M}_{*} \le 9.5$      &$0.11 \pm 0.03 $ &$0.06 \pm 0.02$\\
& $9.5 \le \log_{10} {\mathcal M}_{*} \le 10.0$                                                     &$0.33 \pm 0.03 $ &$0.18 \pm 0.04$\\
& $10.0 \le \log_{10} {\mathcal M}_{*} \le 10.5$                                                    &$0.57 \pm 0.02 $ & $0.51 \pm 0.02$\\       
\hline
$11.0 \le \log_{10} {\mathcal M}_{*} \le 11.5$       &$9.0 \le \log_{10} {\mathcal M}_{*} \le 9.5$     &$0.37 \pm 0.07 $&$0.12 \pm 0.02$\\
& $9.5 \le \log_{10} {\mathcal M}_{*} \le 10.0$                                                   &$0.53 \pm 0.03 $&$0.35 \pm 0.05$\\
& $10.0 \le \log_{10} {\mathcal M}_{*} \le 10.5$                                                  &$0.79 \pm 0.03 $&$0.64 \pm 0.03$\\       
& $10.5 \le \log_{10} {\mathcal M}_{*} \le 11.0$                                                  &$0.90 \pm 0.02 $&$0.75 \pm 0.04$\\                         
\hline
\end{tabular}
\end{center}
\end{table*} 

\begin{itemize} 
 
\item For satellites with hosts in the range $10.0 \le \log_{10}
  {\mathcal M}_{*} \le 10.5$, little change in the red fraction as a
  function of radius is observed, with the red fraction of the
  satellites and small neighbours remaining around the 0.05 level for
  galaxies with $9.0 \le \log_{10} {\mathcal M}_{*} \le 9.5$ and $\sim
  0.14$ for satellites with stellar masses in the range $9.5 \le
  \log_{10} {\mathcal M}_{*} \le 10.0$.

\item For satellites with hosts in the range $10.5 \le \log_{10}
  {\mathcal M}_{*} \le 11.0$, the red fraction increases with satellite
  mass, such that the average red fraction increases from $0.11 \pm 0.03$
  for satellites with masses $9.0 \le \log_{10} {\mathcal M}_{*} \le 9.5$,
  to $0.33 \pm 0.03$ for satellites with masses of $9.5 \le \log_{10}
  {\mathcal M}_{*} \le 10.0$, and $0.57 \pm 0.02$ for satellites with
  masses of $10.0 \le \log_{10} {\mathcal M}_{*} \le 10.5$. It also
  becomes the apparent that the red fraction increases as distance
  decreases. The average red fraction of the satellites is greater
  than the average red fraction of the small neighbours in the same
  mass range. For galaxies with $9.0 \le \log_{10} {\mathcal M}_{*} \le
  9.5$ this difference is $0.0 5 \pm 0.04$, $0.15 \pm 0.04$ for
  galaxies with $9.5 \le \log_{10} {\mathcal M}_{*} \le 10.0$ and $0.06
  \pm 0.04$ for galaxies with $10.0 \le \log_{10} {\mathcal M}_{*} \le
  10.5$.

\item Satellites with host masses $11.0 \le \log_{10} {\mathcal M}_{*} \le
  11.5$ show similar trends and the radial dependence becomes even
  more apparent, especially for the low mass satellites. The average
  red fraction of satellites increases from $0.37 \pm 0.07$ for
  satellites with $9.0 \le \log_{10} {\mathcal M}_{*} \le 9.5$, through to
  $0.90 \pm 0.02$ for the satellites with $10.5 \le \log_{10}
  {\mathcal M}_{*} \le 11.0$. The difference between the red fraction of
  satellites and the small neighbours is observed to be $0.25 \pm
  0.07$ for galaxies in the mass range $9.0 \le \log_{10} {\mathcal M}_{*}
  \le 9.5$, $0.18 \pm 0.07$ for galaxies with $9.5 \le \log_{10}
      {\mathcal M}_{*} \le 10.0$, $0.15 \pm 0.05$ for galaxies with $10.0
      \le \log_{10} {\mathcal M}_{*} \le 10.5$ and $0.15 \pm 0.04$ for
      galaxies with $10.5 \le \log_{10} {\mathcal M}_{*} \le 11.0$.

\end{itemize}

{\revone Our results here support the `galactic conformity' phenomenon,
  whereby red/early-type centrals have a significantly higher fraction
  of red/early-type satellites, noted first by \cite{Weinmann2006} and
  also observed by \cite{Ann2008}.  \cite{Weinmann2006} produce a
  large catalogue of $\sim$ 53\,000 groups, from a sample of $\sim$
  92\,000 galaxies from the SDSS DR2, using the halo-based group
  finder of \cite{Yang2005}. Defining centrals as the brightest group
  members and satellites as the remaining group members, they
  determine the projected halocentric distances (normalised to the
  virial radius of the system $R_{vir}$) of the satellites and halo
  masses of the systems. They find that the early-type fractions of
  satellites increase as a function of increasing luminosity,
  increasing group halo mass and decreasing halocentric distance. For
  halo masses with $14 < \log_{10} {\mathcal M} < 15$, the
  early-type fraction of satellites increases from $\sim 40$ per cent
  at $0.9 R/R_{vir}$ to $\sim$ 60 per cent at $ 0.1 R/R_{vir}$. For
  lower-mass halos with $12 < \log_{10} {\mathcal M} < 13$, the
  early-type fraction of satellites is approximately constant at $\sim$
  25 per cent. This is a similar picture to what we see in
  Figure~\ref{fig:rfhost}}.

In a study more directly comparable to ours, \cite{Ann2008}
investigate satellite systems of isolated hosts, who determine the
morphologies of their satellite and host samples to investigate the
early-type fraction of satellite galaxies as a function of
radius. Splitting their hosts into early and late types, they find
that the early-type satellite fraction for early-type hosts increases
from $\sim 0.2$ at $R_{Proj} = 1$ Mpc to $\sim 0.6$ at $R_{Proj} =
0.05$ kpc, whereas the early-type fraction for late-type hosts remains
roughly constant, $\sim 0.2$ for all projected radii, which is almost
identical to what we find in Figure~\ref{fig:rfhost}(a). By dividing
their sample into hosts with $M_{r} < -20.5$ and $-21.0 < M_{r} <
-19.0$, they show in their figure~2 that the brighter early-type hosts
have an early-type satellite fraction which is 0.05--0.20 greater in
each projected radius bin compared to those with fainter hosts. For
the brighter host bin, the early-type satellite fraction increases
from $\sim 20$ per cent at $R_{Proj} = 1$ Mpc to $\sim 70$ per cent at
$R_{Proj} = 0.05$ Mpc, whereas the early-type satellite fraction for
fainter hosts increases from $\sim 20$ per cent at $R_{Proj} = 1$ Mpc
to $\sim 60$ per cent at $R_{Proj} = 0.05$ Mpc. For the late-type
hosts, the early-type satellite fraction is roughly constant with
projected radius at $\sim 25$ per cent for hosts with $ M_{r} < -20.5$
compared to being roughly constant at $\sim 15$ per cent for hosts
with $-21.0 < M_{r} <-19.0$. Again this trend is similar to our result
that the red fraction of satellites increases with increasing host
mass.

\subsection{Red Fraction as a Function of Mass}

An alternative way of showing how the red fraction varies as a function of
radius can be seen in Figure~\ref{fig:rfmain}(b,d,f), which shows the red
fraction of satellites and small neighbours in mass bins of width $\Delta
\log_{10} {\mathcal M} = 0.25$, for the same host mass ranges as on the
left-hand side. As expected the red fraction of satellites and small
neighbours increases with increasing stellar mass. {\revone
  Figure~\ref{fig:rfmain}(b,d,f) clearly shows that host mass has the biggest
  effect on the red fraction of satellites relative to the small neighbours.}
For the satellites with the lowest host masses ($10.0 \le \log_{10} {\mathcal
  M}_{*} \le 10.5$), there is virtually no difference between the red fraction
of the satellites and small neighbours, confirming that the red fraction has
no radial dependence. For the highest mass hosts, the red fraction of the
satellites is about 0.2 greater than the red fraction of small neighbours.

Using the SDSS DR4 group catalogue of \cite{Yang2007}, \cite{VDBosch2008}
define the most massive members of groups as centrals and the remaining group
members as satellites, producing a sample of 218\,103 centrals and 59\,982
satellites. They then investigate how the red fractions of hosts (centrals)
and satellites by mass compare. In their figure~8, they show that the
difference in the red fraction of centrals and satellites is the greatest at
low masses (the satellites have red fractions which are 20--40 per cent higher
for $\log_{10} {\mathcal M}_{*} \le 10.0 $), converging at higher masses. The
significant red fraction at the lowest satellite masses is discrepant with
what we observe, although this may be explained by the different definitions
of satellite galaxies.

\subsection{Discussion}

We have shown that the red fraction of satellites surrounding isolated
hosts increases as a function of host stellar mass, satellite mass and
decreasing projected separation, indicating the quenching of star
formation in satellite galaxies is more efficient in most massive
systems and acts within $\sim$ 500 kpc of the host. This quenching of
star formation is likely to be due to the removal of gas by a combination of
radiative and gravitational effects of the host, which have been
investigated in previous studies.

One way of shutting off the star formation in satellite galaxies is by
strangulation \citep{Larson1980, Balogh2000}, which is thought to
occur when a satellite's halo is accreted into the larger halo of its
host removing the satellite's hot, diffuse gas and thus causing a slow
decline (on timescales $\tau >$ 1 Gyr) in the star formation as its
supply of cold gas is suppressed, and turning spirals into S0
galaxies. \cite{Kawata2008} simulate the strangulation process,
showing that it acts on satellites within $R < 500$ kpc on timescales
of $\tau \sim 2.0$ Gyr. They conclude the process is more effective in
relaxed groups of galaxies, which are dominated by elliptical galaxies
surrounded by an X-ray emitting intergalactic medium (IGM). Many of
the systems with high mass hosts found in this study could be
described as the system above, making strangulation a likely
explanation for our results. This scenario is also consistent with the
host mass having the largest effect on the red fraction, rather than
satellite mass.

Compared to the slow quenching of star formation which occurs in
strangulation, rapid quenching could occur if enough external pressure
is exerted on the satellite to remove its cold gas, in the process of
ram pressure stripping \citep{Gunn1972,Hester2006}. Thought to turn
spiral galaxies into S0 types, ram pressure stripping is most
effective in clusters and groups of galaxies, where the density of the
hot IGM and velocities of satellites creates sufficient pressure to
overcome the restoring force keeping cold gas in
galaxies. \cite{Hester2006} finds, using simulations of groups, that
low mass satellites are more severely stripped over larger distances,
than larger spiral galaxies which are only stripped at small distances
from the host. {\revone This is not what we see in this study with a
  higher red fraction over 0--0.4\,Mpc, compared to the small
  neighbours, in the highest-mass hosts for all satellite masses 
[see Figure~\ref{fig:rfmain}(e)].} 

Tidal stripping of gas from satellites due to the gravitational effects of the
host is likely to occur simultaneously with ram-pressure
stripping. \cite{Mayer2006}, using N-body simulations of Milky Way sized halos
surrounded by gas rich dwarfs, show that a combination of tidal and ram
pressure stripping is more effective at removing gas than either process
alone. Tidal interactions between satellites and their hosts may explain the
streams of HI gas observed locally as the Magellanic Stream and High Velocity
Clouds \citep{Putman2003} and as the extragalactic HI clouds observed around
M31 \citep{Thilker2004}, M33 \citep{Grossi2008} and the M81/M82 group
\citep{Chynoweth2008}. {\revone Shutting off of star formation might be
  expected to scale with the ratio of satellite to host mass, which is not
  evident in our data.} {\revtwo Consider satellites that are $\sim0.01$--0.1
  times the mass of the host in Figure~\ref{fig:rfmain} [8.5--10 in
    Fig.~\ref{fig:rfmain}(b), 9--10.5 in Fig.~\ref{fig:rfmain}(d), 9.5--11 in
    Fig.~\ref{fig:rfmain}(f)], it is clear that the difference between the
  satellites and small neighbours depends on the host mass but not the ratio
  between the satellite and host mass. This is in agreement with the idea that
  the probability of environment quenching is independent of the satellite
  stellar mass \citep{Peng2011}.}

Another process which is thought to be able to strip galaxies of their
gas is harassment \citep{Moore1996}. This is the heating of a
satellite, occurring when a satellite's DM sub-halo has frequent
high-speed encounters with other DM sub-halos, turning low surface
brightness discs into smaller dwarfs. Although this is usually
considered as a process occurring in clusters containing thousands of
galaxies, $\Lambda$CDM models of the Universe reveal that galaxy systems
are scaled down versions of clusters and should contain thousands of
dark matter sub-halos surrounding galaxies
\citep{Klypin1999,Moore1999}, which means harassment on smaller scales
may be a possibility.

Furthermore, satellite-satellite mergers could occur which would turn
late-type satellites into early types \citep{McIntosh2008}. This
however is unlikely to be a significant way of increasing the red
fraction, due to the low numbers of satellites per host observed.


\section{Conclusions}

We have performed a search for the satellites of isolated galaxies
from the GAMA survey using selection criteria that take into account
stellar mass estimates. Separating galaxies into the red and
blue populations using a colour-mass diagram, we have investigated 
the radial distribution of satellites, and how the red fraction of satellites
varies as function of projected distance, host stellar mass
and satellite stellar mass. 

We find out of a sample of 3\,514 isolated galaxies, 1\,426 which host
a total of 2\,998 satellite galaxies. Separating the galaxies into red
and populations, we find 41.2 per cent of the satellites are blue with
red hosts, 33.3 per cent are red with red hosts, 22.2 per cent are
blue with blue hosts and only 3.3 per cent are red with blue hosts.
We find mean stellar masses and absolute magnitudes of $\log_{10}
{\mathcal M}_{*} = 10.00$ and $M_{r} =-19.96$, and $\log_{10}
{\mathcal M}_{*} = 9.35$ and $M_{r} =-19.30$, for the red and blue
populations of satellites respectively. The average satellite in this
sample is found to be $\sim$ 1/10th of the mass of its host. The mean
$\Delta \log_{10} {\mathcal M}_{*} = 1.18$. Our
main results are as follows:

\begin{enumerate}

\item Parametrizing the projected density of the satellites by
  $\Sigma(R) \propto R^{\alpha}$, we find best fitting slopes of
  $\alpha \simeq -1.0$ for satellites divided into four host stellar
  mass bins (Fig.~\ref{fig:rprojmass}). By dividing the satellite
  hosts into red and blue populations we find the satellites of blue,
  low mass hosts to be more centrally concentrated. Power law slopes
  of $\alpha = -0.94$ and $\alpha = -1.05$ for the satellites of red
  and blue hosts are obtained, respectively. Correcting for
  interlopers the slopes steepen to $\alpha = -1.13$ and $\alpha =
  -1.38$ (Fig.~\ref{fig:rprojcol}).

\item We find there to be a steady increase in the red fraction of
  satellites as projected distance decreases for satellites with hosts
  of stellar mass $\log_{10} {\mathcal M}_{*} \ge 10.5$. The red
  fraction of satellites with hosts of stellar masses $\log_{10}
  {\mathcal M} < 10.5$ shows no trend as a function of radius
  (Fig.~\ref{fig:rfhost}).

\item Sub-dividing the satellite sample into host and satellite mass
  bins reveals that the host mass has the biggest effect on the
  satellite red fraction. Satellites with more massive hosts are more
  likely to be red. Comparing satellites to the small neighbours
  control sample, we find the red fraction is 0.13--0.25 higher for
  satellites of $11.0 \le \log_{10} {\mathcal M}_{*} \le 11.5$ hosts,
  0.06--0.15 higher for $10.5 \le \log_{10} {\mathcal M}_{*} \le 11.0$
  hosts, and almost unchanged for lower-mass $10.0 \le \log_{10}
  {\mathcal M}_{*} \le 10.5$ hosts (Fig.~\ref{fig:rfmain}).
 
\end{enumerate}

{\revtwo The effect of environment, as noted by the difference between the
  satellites and small neighbours, can act over large range in high mass
  haloes and appears to be primarily a function of host mass and not satellite
  mass.}  These results suggest that quenching of star formation by
strangulation is likely to be the main gas removal process in satellite
systems, with other tidal effects, having a smaller contributing effect.
Further study of the radial dependence of the instantaneous star formation
rates of satellite galaxies using H$\alpha$ emission line measurements, is
likely to yield more details about the quenching mechanism.
        
\section*{Acknowledgements}

GAMA is a joint European-Australasian project based around a
spectroscopic campaign using the Anglo-Australian Telescope. The GAMA
input catalogue is based on data taken from the Sloan Digital Sky
Survey and the UKIRT Infrared Deep Sky Survey. Complementary imaging
of the GAMA regions is being obtained by a number of independent
survey programs including GALEX MIS, VST KIDS, VISTA VIKING, WISE,
Herschel-ATLAS, GMRT and ASKAP providing UV to radio coverage. GAMA is
funded by the STFC (UK), the ARC (Australia), the AAO, and the
participating institutions. The GAMA website is
http://www.gama-survey.org/. MP acknowledges STFC for a postgraduate
studentship. IKB and PAJ acknowledge STFC for funding. We acknowledge
the IDL Astronomy User's Library, and IDL code maintained by
D.~Schlegel (IDLUTILS) as valuable resources.

We thank the anonymous referee for useful comments which have improved
the content, clarity and presentation of this paper.

\bibliographystyle{mn2e}
\bibliography{ALLPAPERSSATS}

\bsp
\label{lastpage}

\end{document}